\documentclass[12pt]{JHEP} 
\usepackage{epsfig}
\newcommand\fverb{\setbox\pippobox=\hbox\bgroup\verb}
\newcommand\fverbdo{\egroup\medskip\noindent%
			\fbox{\unhbox\pippobox}\ }
\newcommand\fverbit{\egroup\item[\fbox{\unhbox\pippobox}]}
\newbox\pippobox

\def\del{\partial}

\def\gap#1{\vspace{#1 ex}}

\def\half{\frac{1}{2}}

\def\H{{\cal H}}
\def\F{{\cal F}}

\def\psid{\psi^\dagger}
\def\phid{\phi^\dagger}
\def\ad{a^\dagger}

\def\eps{{\epsilon}}

\def\myitem#1{\gap2\noindent\underbar{#1}\gap2}

\def\be{\begin{equation}}
\def\ee{\end{equation}}
\def\ba{\begin{array}{l}}
\def\ea{\end{array}}
\def\bea{\begin{eqnarray}}
\def\eea{\end{eqnarray}}
\def\beas{\begin{eqnarray*}}
\def\eeas{\end{eqnarray*}}
\def\eq#1{(\ref{#1})}

\def\nn{\nonumber\\}

\def\ket#1{| #1 \rangle}
\def\bra#1{ \langle #1 |}
\def\overlap#1#2{\langle #1 | #2 \rangle}

\def\vev#1{\langle #1 \rangle}

\def\mattwo#1#2#3#4{\left(\begin{array}{cc}#1&#2\\#3&#4\end{array}\right)}

\def\winf{\ensuremath{W_\infty}}
\def\ads{$AdS_5 \times S^5$}

\title{Exact operator bosonization of finite number
of fermions in one space dimension}
\author{Avinash Dhar$^{1}$, Gautam Mandal$^{1}$
and  Nemani V Suryanarayana$^{2}$ \\
$^{1}$ Tata Institute of Fundamental Research, Homi Bhabha Road,\\
Mumbai 400 005, India, \\
$^{2}$ Perimeter Institute of Theoretical Physics,
31 Caroline Street North\\ 
Ontario, Canada N2L 2Y5.
\\~~\\
\email{adhar@theory.tifr.res.in,mandal@theory.tifr.res.in, 
vnemani@perimeterinstitute.ca
}}

\preprint{TIFR/TH/05-36 
}

\abstract{We derive an exact operator bosonization
of a finite number of fermions in one space dimension. The fermions
can be interacting or noninteracting and can have an arbitrary
hamiltonian, as long as there is a countable basis of states in the
Hilbert space. In the bosonized theory the finiteness of the number of
fermions appears as an ultraviolet cut-off. We discuss implications
of this for the bosonized theory. We also discuss applications of 
our bosonization to one-dimensional fermion systems dual to 
(sectors of) string theory such as LLM geometries and c=1 matrix model.}


\begin{document}


\section{Introduction}

The recent study by Lin, Lunin and Maldacena \cite{LLM} of a class of
half-BPS type IIB geometries in asymptotically \ads~spaces, offers an
excellent laboratory to explore aspects of quantum gravity.  In the
boundary super Yang-Mills theory, the corresponding half-BPS states
are described by $N$ free fermions in a harmonic potential
\cite{Corley:2001zk,Berenstein:2004kk,Takayama:2005yq}. At large
$N$, there is a semiclassical description of the states of this 
system in terms of droplets of fermi fluid in phase space; LLM showed
that there is a similar structure in the classical geometries in the
bulk. The semiclassical correspondence is already remarkable in the
sense that it exhibits a noncommutative structure of two of the space
coordinates \cite{LLM,Mandal:2005wv}; however, finite $N$ effects,
corresponding to fully quantum mechanical aspects of bulk gravity,
open up more interesting questions
\cite{Dhar:2005qh}. While it has been shown that
semiclassically small fluctuations of the droplet boundaries
correspond to small gravitational fluctuations around the classical
geometries \cite{LLM,Grant:2005qc,Maoz:2005nk}, at finite $N$  
only those fluctuations of the fermi system which have low enough
excitation energy compared to $N$ can be identified with gravity modes
propagating in the bulk (we will elaborate on this in Section
\ref{fuzzy.sec}).
Excitations with energy comparable to or higher than $N$ do not
correspond to gravitons but to non-local objects in the bulk, namely
giant gravitons or dual giant gravitons
\cite{McGreevy:2000cw,Grisaru:2000zn,Hashimoto:2000zp}. Even more  
remarkable is the fact that the fermi partition function can be 
mapped onto the partition function for giant gravitons or dual 
giant gravitons \cite{Suryanarayana:2004ig} alone, without involving 
any low-energy gravitational degrees of freedom at all. This seems to
suggest that, at least in the half-BPS sector,  the bulk geometry 
has a nontrivial structure at a small enough length scale whose 
precise value depends on $N$, and that below this length scale 
gravitational phenomena in the bulk are 
described by degrees of freedom that are quite different from 
the degrees of freedom that characterize low-energy gravitational 
fluctuations. In such an interpretation, low-energy gravity modes 
would be ``composites'' of the microscopic degrees of freedom. 
This provides a motivation to look for an exact bosonization of the 
finite $N$ fermi system, which should provide the ``right'' 
variables to describe bulk gravity consistent with such a structure.

Bosonization of a system of finite number of fermions is an
interesting problem in its own right, with many potential applications
in quantum field theory as well as in condensed matter theory. For
this reason the problem has received attention for more than 
half a century now. Approximate solutions have long been found in case of
free nonrelativistic fermions near the fermi level \cite{Tomonaga}
where the fermion density turns out to be the spatial derivative of
the bosonic field. This bosonization becomes exact when the fermions
are relativistic and are infinite in number
\cite{Coleman,Mandelstam}. In the case of free nonrelativistic 
fermions in an inverted harmonic oscillator potential in one space
dimension, arising in the context of $c=1$ matrix models, approximate
bosonization in terms of fermion density gives rise to the massless
boson of two-dimensional noncritical string theory (usually called the
``tachyon'')
\cite{Sengupta-Wadia,Das-Jevicki,Gross-Klebanov,Polchinski}. 
An exact bosonization of fermions in an arbitrary potential in one 
space dimension was found in \cite{DMW-nonrel,DMW-path}
in terms of the Wigner phase space density of the fermions, but the
bosonic variable satisfies an infinite dimensional nonabelian algebra
(W-infinity), instead of the Heisenberg algebra, and it satisfies a
quadratic constraint; the approximate bosonization in terms of the
position space density can be derived from it. Recently it has been
noticed \cite{Corley:2001zk,Berenstein:2004kk,Itzhaki:2004te,
LLM,Suryanarayana:2004ig}  
that the spectrum, and consequently the partition function, of $N$
nonrelativistic fermions in a harmonic oscillator potential
(discussed in the previous paragraph) agrees
with that of a system of free nonrelativistic bosons which are
infinite in number but each of which moves in an equally spaced
$N$-level system (similar to harmonic oscillator energy levels with an
upper cutoff). Indeed the fermionic spectrum also agrees 
\cite{Suryanarayana:2004ig} with
that of a second bosonic system with a finite number $N$ of particles,
each moving in a harmonic oscillator potential. These two bosonic
spectra represent those of giant gravitons \cite{McGreevy:2000cw} and
dual giant gravitons \cite{Grisaru:2000zn,Hashimoto:2000zp}
respectively.

In the present work, we derive an {\it exact operator equivalence}
between a system of $N$ fermions in one space dimension and two
different bosonic systems satisfying the usual commutation relations
(Heisenberg algebra). The two bosonic systems are reminiscent of giant
gravitons and dual giant gravitons, but they appear in bosonization of
fermions moving in any potential. Perhaps the most remarkable effect
of having a finite number of fermions is that it gives rise to
fuzziness in coordinate space in the bosonized theory. This fuzziness
can be seen directly in the bosonized theory and arises because $N$
provides a high energy cut-off on the basic bosonic degrees of
freedom. It can also be seen by reformulating the bosonized theory as
a theory on a lattice with spacing given by $1/N$. In the LLM context,
what this means is that for finite $N$ we have a direct derivation of
the appearance of a short-distance cut-off in the bulk string/gravity
theory, at least in the half-BPS sector. This is consistent with what
we expect from stringy exclusion principle
\cite{chp3:MalStr98}. Earlier works on the appearance of graininess on
the gravity side of AdS/CFT correspondence are
\cite{Jevicki:1998rr,Jevicki:1998bm,Ho:1999bn,Jevicki:2000it}.

This paper is organized as follows. In the next section, we first give
the rules for our first bosonization which maps the system of $N$
fermions to a system of bosons each of which can occupy a state in an
$N$-dimensional Hilbert space $\H_N$. The finiteness of
the number of fermions reappears in the bosonized theory as finite
dimensionality of the single-particle Hilbert space. Consequences of
this for the bosonized theory are discussed in Section
\ref{prop}. In particular, we argue that the quantum phase space of
the bosons is fuzzy and compact. Actually finite $N$ is responsible
for graininess even in coordinate space. This is seen more directly in
a lattice formulation of the bosonized theory, with lattice spacing
$1/N$, which is also discussed in this section. Section \ref{prop} also
includes a detailed discussion of the bosonic phase space density. In
the LLM context, the bosonic density as a function of the phase space
has the appearance of a rugged circular ``cake'' with an approximately
fixed diameter, confirming the interpretation of the bosons as giant
gravitons. This is to be contrasted with the fermionic phase space
density which looks like droplet configurations. Section \ref{llm} is
devoted to a detailed discussion of this and other aspects of
application of our bosonization in the LLM context. In particular, we
argue that at finite $N$, the LLM gravitons must be fuzzy in an
essential way. Section \ref{second} describes the second bosonization.
The essential difference with the first bosonization is that the
number $N$ now appears as an upper limit on the total number of
bosons. Application of our bosonization to the $c=1$ matrix model 
is briefly sketched  in Section \ref{c=1}.  Some
details of computations are given in Appendices
\ref{details} and \ref{husimi.sec}. In Appendix \ref{rep} we discuss
an important byproduct of our bosonization, namely the bosonization of
$N$ fermions in a finite ($K$) level system and the resulting bosonic
construction of representations of $U(K)$,
which is different from the well-known Schwinger
representation.

\section{\label{bose}The First Bosonization}

Let us first set up the notation.
Consider a system of $N$ fermions each of which can occupy
a state in an infinite-dimensional Hilbert space  $\H_f$.
Suppose there is a countable basis of $\H_f:
\{ \ket{m}, m=0,1, \cdots, \infty\}$. For example, this could
be the eigenbasis of a single-particle hamiltonian, $\hat h \ket{m} =
{\cal E}(m) \ket{m}$, although other choices of basis would do equally
well, as long as it is a countable basis.  In the second quantized
notation we introduce creation (annihilation) operators
$\psi^\dagger_m$ ($\psi_m$) which create (destroy) particles in the
state $\ket{m}$. These satisfy the anticommutation relations
\be
\{ \psi_m, \psid_n\}= \delta_{mn}
\label{anticom}
\ee
The $N$-fermion states are  given by (linear combinations of)
\be
\ket{f_1, \cdots, f_N}= \psid_{f_1}\psid_{f_2} \cdots
\psid_{f_N} \ket{0}_F
\label{fermi-state}
\ee
where $f_m$ are arbitrary integers satisfying 
$0\le f_1 < f_2 < \cdots < f_N$, and 
$\ket{0}_F$ is the usual 
Fock vacuum annihilated by $\psi_m, m=0,1, \cdots, \infty$.

One can create any of the states $\ket{f_1, \cdots, f_N}$ from the
state $\ket{F_0} \equiv \ket{0,1, \cdots, N-1}$ by repeated
application of operators
\be
\Phi_{mn} = \psid_m \, \psi_n
\label{phi-mn}
\ee
Properties of $\Phi_{mn}$ and related operators, including the
Wigner and Husimi distributions, are listed in Appendix B.

We will map the above fermionic system to a system of bosons each of
which can occupy a state in an $N$-dimensional Hilbert space
$\H_N$. Suppose we choose a basis $\{\ket{i},~i=1,\cdots,N\}$ of
$\H_N$. In the second quantized notation we introduce creation
(annihilation) operators $a^\dagger_i$ ($a_i$) which create (destroy)
particles in the state $\ket{i}$. These satisfy the commutation
relations
\be
[a_i, a^\dagger_j]= \delta_{ij}, \quad i,j=1, \cdots, N
\label{oscillator}
\ee 
A state of this bosonic system is given by (a linear combination
of)
\be
\ket{r_1, \cdots, r_N}= {(a_1^\dagger)^{r_1}\cdots
(a_N^\dagger)^{r_N} \over \sqrt{r_1 ! \cdots r_N!}} \ket{0}_B
\label{bose-state}
\ee

\subsection{The bosonization formulae}

The bosonization formulae are written most economically using 
the notions of operator delta and theta-functions. These are 
defined for any operator $\hat O$ as follows:
\be
\delta (\hat O) \equiv 
\int_0^{2\pi} \frac{d\theta}{2\pi}~\exp(i \theta \hat O),
\label{delta}  
\ee
and
\be
\theta_+(\hat O) \equiv \sum_{m=0}^\infty \delta(\hat O-m), \quad
\theta_-(\hat O) \equiv 1 - \theta_+(\hat O).
\label{theta}
\ee 
Furthermore, we will also need to introduce the following operators:
\be
\sigma_k \equiv \frac{1}{\sqrt{\ad_k a_k + 1}} a_k, \quad
\sigma^\dagger_k \equiv \ad_k\frac{1}{\sqrt{\ad_k a_k + 1}}
\label{sigmas}
\ee
These operators have interesting properties. In particular, the
following relations are useful in obtaining the bosonization formulae
given below. 
\be
\sigma_k~\sigma^\dagger_k=1, \quad 
\sigma^\dagger_k~\sigma_k=\theta_+ (\ad_k a_k-1). 
\label{sigmarelations}
\ee

Since delta-function operators of the form $\delta (\sum_{i=1}^n
a^\dagger_{k_i} a_{k_i} - p)$ will appear quite often in the formulae
given below, we give an alternative, more familiar, expression for it
in terms of more elementary operators. First recall
the following representation of the
operator $|m\rangle \langle n|$ in a harmonic oscillator Hilbert
space with raising (lowering) operators $\ad (a)$:
\begin{equation}
|m \rangle \langle n| = \frac{1}{\sqrt{m!n!}} :a^{\dagger\, m}
 e^{-a^\dagger a} a^n:
\end{equation}
where $:\,:$ represents normal ordering. This implies 
\begin{equation}
\delta (a^\dagger a - n) = |n\rangle \langle n| =
 \frac{1}{n!} :a^{\dagger\, n} 
 e^{-a^\dagger a} a^n: 
\end{equation}
which can be used to write
\begin{eqnarray}
\delta (\sum_{i=1}^n a^\dagger_{k_i} a_{k_i}-p) &=&
\sum_{r_{k_1}+\cdots + r_{k_n} = p} \prod_{i=1}^n \delta
(a_{k_i}^\dagger a_{k_i} - r_{k_i}) \cr
&=&
\sum_{r_{k_1}+\cdots + r_{k_n} = p} \frac{1}{r_{k_1}! \cdots r_{k_n}!}
: a^{\dagger \, r_{k_1}}_{k_1} \cdots a^{\dagger \, r_{k_n}}_{k_n} \, 
e^{- \sum_{i=1}^n a^\dagger_{k_i} a_{k_i}} 
a^{r_{k_1}}_{k_1} \cdots a^{r_{k_n}}_{k_n}: \cr
&&
\end{eqnarray}
Similarly, one can give alternative definitions of the operators
$\delta(\sum_{m=m_1}^{m_2} \psi_m^\dagger \psi_m)$ which appear
extensively in Eqns.(\ref{fermionize1},\,
\ref{fermionize2}) below. We first note the identities
\begin{equation}
\delta(\psi_n^\dagger \psi_n) = 1-\psi_n^\dagger \psi_n, 
~~~ \delta
(\psi_n^\dagger \psi_n -1) = \psi_n^\dagger \psi_n.
\end{equation}
These identities enable us to write the following alternative
expressions for some fermionic delta-functions:
\begin{equation}
\delta(\sum_{m=m_1}^{m_2} \psi_m^\dagger \psi_m)= \prod_{m=m_1}^{m_2}
\delta (\psi_m^\dagger \psi_m) = \prod_{m=m_1}^{m_2} (1-\psi_m^\dagger
\psi_m). 
\end{equation}

We are now ready to describe the bosonization formulae.

\subsubsection{Fermionic representation of bosonic oscillators}
 
We will first define the bosonic creation and annihilation 
operators in terms of their action on the states
of the fermion system. We have, on a general $N$-fermion state 
(\ref{fermi-state}),
\bea
\ad_k~\ket{f_1, \cdots, f_N} = && \sqrt{f_{N-k+1}-f_{N-k}}~
\ket{f_1, \cdots, f_{N-k}, f_{N-k+1}+1, \cdots, f_N+1}, 
\, k=1,...,N-1
\nn
\ad_N~\ket{f_1, \cdots, f_N} = && \sqrt{f_1+1}~\ket{f_1+1, \cdots, 
f_N+1}.
\label{boson1}
\eea
Thus, $\ad_k$ moves each of the top $k$
fermions, counting down from  
the topmost filled level, up by one step. Similarly, the action of 
$a_k$ is to move each of the top $k$ fermions down by one step:
\bea
a_k~\ket{f_1, \cdots, f_N} = && \sqrt{f_{N-k+1}-f_{N-k}-1}~
\ket{f_1, \cdots, f_{N-k}, f_{N-k+1}-1, \cdots, f_N-1}, 
\, k=1,...,N-1
\nn
a_N~\ket{f_1, \cdots, f_N} = && \sqrt{f_1}~\ket{f_1-1, \cdots,
f_N-1}. 
\label{boson2}
\eea 
The reasoning that led us to these
expressions for the bosonic 
creation and annihilation operators has been explained in Appendix
\ref{origin}.
Here we simply mention that these definitions of the bosonic 
creation and annihilation operators satisfy the oscillator algebra 
\eq{oscillator}. The reader can find a proof of this in Appendix 
\ref{algebraproof}.
Note that the state $\ket{F_0} = \ket{0,\, 1, \,\cdots,
N-1}$ is special since it is annihilated by all the annihilation
operators $a_k,~k=1, \, 2, \, \cdots, \, N$. Therefore, as expected, 
$\ket{F_0}$ is the oscillator vacuum state $\ket{0}_B$.

One can also give operator expressions for the oscillator creation
and annihilation operators in terms of the fermion bilinears. We have,
\bea
\ad_k \equiv && \sum_{m_k> m_{k-1} > \cdots > m_0}
\sqrt{m_1 - m_0}~(\psid_{m_0}\psi_{m_0})(\psid_{m_1 + 1} \psi_{m_1}) \cdots
(\psid_{m_k + 1} \psi_{m_k}) \nn
&& ~~\times \delta \biggl (\sum_{m=m_0 +1}^{m_1 -1} 
\psid_m \psi_m \biggr )~
\delta \biggl (\sum_{m=m_1 +1}^{m_2 -1}\psid_m \psi_m \biggr ) \cdots
~\delta \biggl (\sum_{m=m_{k-1} +1}^{m_k -1} \psid_m \psi_m \biggr ) \nn
&& ~~\times 
\delta \biggl (\sum_{m=m_k +1}^{\infty}\psid_m \psi_m \biggr ), 
\quad \quad \quad \quad k=1, 2, \cdots , (N-1)
\label{fermionize1}
\eea
and
\bea
\ad_N \equiv && \sum_{m_N> m_{N-1} > \cdots > m_1}
\sqrt{m_1 + 1}~(\psid_{m_1 + 1} \psi_{m_1}) \cdots 
(\psid_{m_N + 1} \psi_{m_N}) \nn
&& ~~~~~~~~~~~~\times 
\delta \biggl (\sum_{m=m_1 +1}^{m_2 -1}\psid_m \psi_m \biggr ) \cdots
~\delta \biggl (\sum_{m=m_{N-1} +1}^{m_N -1} \psid_m \psi_m \biggr ) \nn
&& ~~~~~~~~~~~~\times 
\delta \biggl (\sum_{m=m_N +1}^{\infty}\psid_m \psi_m \biggr ).
\label{fermionize2}
\eea
The annihilation operators are obtained from these by conjugation.
These expressions look complicated, but it is easy to see that when
acting on a generic fermion state \eq{fermi-state}, because of the
operator delta-functions, only that term in the sum survives for which
$m_k=f_N, \, m_{k-1}=f_{N-1},\cdots,\, m_0=f_{N-k}$ in
\eq{fermionize1} and $m_N=f_N, \, m_{N-1}=f_{N-1},\cdots,\, m_1=f_1$
in \eq{fermionize2}.  This reproduces \eq{boson1}. One can similarly
reproduce \eq{boson2} from the action of conjugates of
\eq{fermionize1} and \eq{fermionize2} on the generic fermion state
\eq{fermi-state}.

Note that the bosonic oscillators are mapped into combinations of
fermion {\it bilinears}, \eq{phi-mn}. This is because we are
interested in a fixed fermion number sector of the fermion Fock space.

\subsubsection{Fermion bilinears $\psid_m\psi_n$ in terms
of bosonic oscillators}
 
The inverse map gives fermions in terms of the bosonic oscillators.
Since the total number of fermions is conserved, this bosonization map
can only relate fermion bilinears to the bosonic operators. The
generic fermion bilinear is $\psid_m \, \psi_n$, where $m, \,n = 0, \,1,
\cdots , \,\infty$, but it is sufficient for us to obtain a bosonized
expression for the bilinear for $m \geq n$ only. The bosonized
expression for $m < n$ can be obtained from this by conjugation.
Before going to the most general expression (see \eq{bosonizeapp}) we
will first describe the relatively simple expressions obtainable for (i)
for small values of $(m-n)$ and arbitrary $N$, and (ii) small
values of $N$ and $(m-n)$ any positive integer. We list below
expressions for a few examples of both kinds.

\myitem{$N=1$}: 

This is the simplest case. Here there is only one creation
(annihilation) operator, $\ad$ ($a$). We have,
\bea
\psid_n~\psi_n = && \delta (\ad a - n ) \nn
\psid_{n+m}~\psi_n = && {\sigma^\dagger}^m~\delta (\ad a - n ) 
\label{bosonizeN1}
\eea

\myitem{$N=2$}:

In this case there are two creation (annihilation) operators, $\ad_1$
($a_1$) and $\ad_2$ ($a_2$). The bosonized expressions are now more
complicated than for $N=1$ case, but still manageable. We have,
\bea
\psid_n~\psi_n = && \delta (\ad_1 a_1 + \ad_2 a_2 - n +1)
+ \delta(\ad_2 a_2 - n ) \nn
\psid_{n+m}~\psi_n = && {\sigma^\dagger_1}^m~ 
\delta (\ad_1 a_1 + \ad_2 a_2 - n + 1 ) + 
\sigma_1^m~{\sigma^\dagger_2}^m~\theta_+ (\ad_1 a_1 - m)~
\delta (\ad_2 a_2 - n ) \nn
&& - \sum_{r_1=0}^{m-2} {\sigma^\dagger_1}^{m-2-r_1}~\sigma_1^{r_1}~
{\sigma_2^\dagger}^{r_1+1}~\delta (\ad_1 a_1 - r_1 )~\delta (\ad_2 a_2 - n )
\label{bosonizeN2}
\eea

\myitem{Arbitrary $N$}: 

In this case relatively simple expressions exist only for small values  
of $(m-n)$. We give below expressions for $m=n, \, n+1$ and $n+2$.
\bea 
\psid_n~\psi_n = &&
\sum_{k=1}^N \delta\biggl (\sum_{i=k}^N \ad_i a_i-n+N-k \biggr)  
\nn
\psid_{n+1}~\psi_n = && \sigma_1^\dagger~ \delta \biggl (\sum_{i=1}^N
\ad_i a_i-n+N-1 \biggr) \nn && \!\!\!\!\! + \sum_{k=1}^{N-1}
\sigma_k~\sigma^\dagger_{k+1}~\theta_+(\ad_k a_k-1)~ \delta \biggl
(\sum_{i=k+1}^N \ad_i a_i-n+N-k-1 \biggr) 
\nn 
\psid_{n+2}~\psi_n = &&
{\sigma_1^\dagger}^2~ \delta \biggl (\sum_{i=1}^N \ad_i a_i-n+N-1
\biggr) \nn && \!\!\!\!\!\!\! +\sum_{k=1}^{N-1}
\sigma_k^2~{\sigma_{k+1}^\dagger}^2~\theta_+(\ad_k a_k-2)~ \delta
\biggl (\sum_{i=k+1}^N \ad_i a_i-n+N-k-1 \biggr) \nn && \!\!\!\!\!\!\! 
-\sum_{k=2}^{N-1} \sigma_{k-1}~\sigma^\dagger_{k+1}~
\theta_+(\ad_{k-1} a_{k-1}-1)~ \delta (\ad_k a_k)~\delta \biggl
(\sum_{i=k+1}^N \ad_i a_i-n+N-k-1 \biggr) \nn && \!\!\!\!\!\!\!
-\sigma_2^\dagger~\delta (\ad_1 a_1)~ \delta \biggl (\sum_{i=1}^N
\ad_i a_i-n+N-2 \biggr) 
\label{bosonizetxt} 
\eea
Bosonized expression for the generic bilinear for arbitrary $N$ is
rather complicated and not particularly illuminating. It has therefore
been relegated to the Appendix and is given in \eq{bosonizeapp}.

To check the validity of the bosonization formulae, we need to check
that the (\winf) algebra \eq{winf}, expressing the commutation
relation of fermion bilinears, works out. The \winf\ algebra \eq{winf}
works out for $N=1,~2$ if one uses the bosonized expressions
\eq{bosonizeN1} and \eq{bosonizeN2}. The first case is somewhat
trivial. However, the second case is nontrivial and the satisfaction
of the algebra \eq{winf} requires delicate cancellations, as we have
shown in Appendix A. We have not yet completed a check of the algebra
for general $m,~n$ in the case of arbitrary $N$ because of the
complexity of the relevant formula, \eq{bosonizeapp}. However, because
of the nontrivial way in which it checks out for $N=2$, we are
confident that it will also check out for arbitrary $N$. Moreover, as
proved in Appendix A, the algebra works out for arbitrary $N$ for
small values of $(m-n)$.

\subsubsection{The bosonized hamiltonian}

Let us now discuss the bozonization of the hamiltonian.
Before discussing the most generic case \eq{interacting}, let
us first ignore the fermion-fermion interactions. Let 
${\cal E}(m),~m=0, 1, 2, \cdots$ be the exact single-particle 
spectrum of the fermions in that case (${\cal E}(m)$
are the eigenvalues of the matrix $E_{mn}$ 
of \eq{interacting}). Then, the hamiltonian is given by
\be
H = \sum_{m=0}^\infty {\cal E}(m) ~ \psid_m \psi_m
\label{free-ham}
\ee
Using the bosonization formula \eq{bosonizetxt}, a bosonized 
expression for the hamiltonian can be easily worked out: 
\bea
H &=& \sum_{m=0}^\infty {\cal E}(m) 
\sum_{k=1}^N \delta\biggl (\sum_{i=k}^N \ad_i a_i-m+N-k \biggr) \nn
&=& \sum_{k=1}^N {\cal E} \biggl (N-k+\sum_{i=k}^N \ad_i a_i \biggr ).
\label{hamiltonian}
\eea
The first equality above follows from the first line 
of \eq{bosonizetxt}.
Notice that, in general, the hamiltonian will not be quadratic in 
the bosonic creation and annihilation operators. Nevertheless, the 
bosonic states \eq{bose-state} diagonalize the hamiltonian because
it only involves the occupation number operators. 

For most potentials, the exact spectrum ${\cal E}(m)$
of the single-particle hamiltonian is unlikely to be known. 
However, what we need for the purposes of bosonization is any 
countable basis, which could be provided, for example, by a 
part of the single-particle 
hamiltonian $\hat h$ that is exactly diagonalizable. Thus, suppose, 
$\hat h=\hat h_0+\hat h_1$ such that $\ket{m},~m=0, 1, 2,~ \cdots$ 
is a countable eigenbasis of $\hat h_0$ with eigenvalues 
${\cal E}_0(m)$. Note that we do not require $\hat h_1$ to be small
compared to $\hat h_0$. Then, we have
\bea
H &=& \sum_{m,n=0}^\infty \biggl ({\cal E}_0(m) \delta_{mn}+
<m|\hat h_1|n> \biggr )~\psid_m~\psi_n \nn
&=& \sum_{k=1}^N {\cal E}_0 \biggl (N-k+\sum_{i=k}^N \ad_i a_i \biggr )
+\sum_{m,n=0}^\infty <m|\hat h_1|n>~\psid_m~\psi_n.
\label{hamiltonian2}
\eea
Given $\hat h_1$, the matrix elements $<m|\hat h_1|n>$ can be easily
calculated. Typically these matrix elements will not be diagonal 
\footnote{The allowed values of $m,~n$ depend on the system under
consideration. For example, let $\hat h_0$ be the harmonic oscillator 
hamiltonian and $\hat h_1$ a quartic anharmonic piece. In this case, 
the matrix element $<m|\hat h_1|n>$ vanishes unless $|m-n| \leq 4$.}. 
Thus, to obtain the bosonized form
of the hamiltonian, we will need to use not only the bosonized
expression for the fermion bilinear in the first of \eq{bosonizetxt},
but also the expression \eq{bosonizeapp} for the general bilinear
$\psid_m \, \psi_n$. It follows that in this basis
the hamiltonian \eq{hamiltonian2} is not automatically diagonal. 
The ``law of conservation of difficulty'' is operative here; the 
problem of finding the exact fermionic single-particle spectrum 
has reappeared as the problem of diagonalizing this bosonic 
hamiltonian! 

Let us now consider a few special cases in some detail to illustrate
how our bosonization works in practice.

\myitem{Fermions in a harmonic potential}

A drastic simplification occurs in this case since this potential
gives rise to an equally spaced spectrum, namely ${\cal
E}(m)=c_1~m+c_2$. In this case, an exact expression for the bosonized
hamiltonian can be worked out and it corresponds to a bunch of simple
harmonic oscillators (see \eq{hamiltonian}):
\be
H_{\rm equal-spacing}=c_1 \sum_{k=1}^N k~\ad_k a_k + H_{\rm vac},
\label{equal-spacing}
\ee 
where $H_{\rm vac}=\frac{c_1}{2}N(N-1)+c_2 N$ is the energy of the 
fermi ground state.

\myitem{Free fermions on a circle}

In this case the single-particle spectrum is given by ${\cal
E}(n)=cn^2, ~c=2 \pi^2 \hbar^2/mL^2$, where $m$ is the mass
of a fermion and $L$ is the circumference of the circle. The novelty
here is that except for the ground state, each of the levels is doubly
degenerate.  We will consider this case in some detail since it
illustrates the generality of our bosonization. Moreover,
this example is among the rare exactly solvable cases with a nonlinear
spectrum.

The normalized single-particle eigenstates (assuming periodic boundary
conditions) are $\chi_{\pm n}(x)=\frac{1}{\sqrt L}~e^{\pm i 2\pi n
x/L}, ~n=0,\,1,\,2,\,\cdots$ ($\chi_0(x) =\frac{1}{\sqrt L}$ is
non-degenerate). The mode expansion of the fermion field will involve
the corresponding annihilation (creation) operators $\tilde
\psi_{\pm n}$ ($\tilde \psid_{\pm n}$). To make contact with our
bosonization, we introduce the following identifications:
$\psi_{2n-1}=\tilde \psi_{+n}$ and $\psi_{2n}=\tilde \psi_{-n}$ for
$n=1,\,2,\,\cdots$ ($\psi_0$ corresponds to the
constant mode $\chi_0(x)$). This identification maps the two fermion modes
corresponding to each of the degenerate single-particle levels to two
consecutive modes in an auxiliary fermion problem which our
bosonization technique can handle. Using this mapping we can now
transcribe all the bosonization formulae to this case. In particular,
the bosonized hamiltonian can be obtained as follows:
\bea
H_{\rm circle} &=& c \sum_{n=1}^\infty n^2 \biggl 
[\tilde \psid_{+n} \, \tilde \psi_{+n}+\tilde \psid_{-n} \, \tilde
\psi_{-n} \biggr ] \nn
&=& c \sum_{k=1}^N \sum_{n=1}^\infty n^2 \biggl
[\psid_{2n-1} \, \psi_{2n-1}+\psid_{2n} \, \psi_{2n} \biggr ] \nn
&=& c \sum_{k=1}^N \sum_{n=1}^\infty n^2 \biggl
[\delta \biggl ( \sum_{i=k}^N \ad_i a_i-2n+1+N-k \biggr )
+\delta \biggl ( \sum_{i=k}^N \ad_i a_i-2n+N-k \biggr ) \biggr ] \nn
&=& \frac{c}{4} \sum_{k=1}^N \biggl
[N-k+\sum_{i=k}^N \ad_i a_i + 
\frac{1}{2} \biggl ( 1-(-1)^{N-k+\sum_{i=k}^N \ad_i a_i} \biggr ) 
\biggr ]^2 
\label{circle}
\eea
The second equality above follows from our mapping of the degenerate
levels to odd and even level of the auxiliary fermion problem
and the third follows from the first of the bosonization formulae
in \eq{bosonizetxt}. As an example, let us compute the energy of the
vacuum state using the bosonized hamiltonian. On the vacuum state, 
the oscillator term vanishes. We get
\be
H_{\rm circle}~\ket{0}_B=\frac{c}{4} \sum_{k=1}^N \biggl
[N-k+\frac{1}{2} \biggl ( 1-(-1)^{N-k} \biggr ) \biggr ]^2~\ket{0}_B.
\ee
For $N$ even, the eigenvalue becomes
$2c[1^2+2^2+\cdots+(N/2-1)^2]+c(N/2)^2$, while for $N$ odd we get
$2c[1^2+2^2+\cdots+{(\frac{N-1}{2}})^2]$. These are precisely the
correct energy eigenvalues of the $N$-fermion ground state in the two
cases. One can similarly check that the bosonized hamiltonian in 
\eq{circle} correctly gives the energy eigenvalues for excited states.

For small fluctuations around the Fermi vacuum, one
can define a semiclassical limit in which the standard
relativistic boson emerges as low energy excitations.
These excitations coincide with the ones created by
the $\ad$-oscillators. Details of this calculation
will be presented elsewhere.

Notice that the fermionic hamiltonian (first line of \eq{circle}) is
manifestly invariant under $n \rightarrow -n$, for any given $n$. It
is reflected as degeneracies in the spectrum. This manifest symmetry
of the hamiltonian is lost in the bosonized form (last line of
\eq{circle}), although of course the bosonic spectrum does display the
appropriate degeneracies. For example, consider $N$ even. In this case
the fermi ground state is doubly degenerate since the single fermion
at the top can occupy either of the two degenerate states labeled by
$n=\pm N/2$. In the bosonic language, this pair of degenerate states
is $\ket{0}_B$ and $\ad_1\ket{0}_B$. Similarly, for $N$ odd, the fermi
ground state is unique, but the first excited state has four-fold
degeneracy. The corresponding four degenerate bosonic states are
$\ad_1\ket{0}_B,~{\ad_1}^2\ket{0}_B,~\ad_2\ket{0}_B$ and
$\ad_2~\ad_1\ket{0}_B$. There will be a maximum of $2^N$ degenerate
states in the most general case. It is an important problem to
understand in a more systematic way this realization of the symmetry
structure in the bosonic theory. This would open up the possibility of
applications of our bosonization to problems in higher than $1$ space
dimension. For example, in $3$ space dimensions in a potential with
spherical symmetry, one can proceed as above and give some assignment
of degenerate angular momentum states to the auxiliary fermion problem
\footnote{One could choose an assignment by switching on a small
magnetic field that breaks rotational symmetry. The original problem
is recovered by letting the magnetic field go to zero after
bosonization is done.}. The important issue would then be to
understand how rotational symmetry is realized in the bosonized
theory. A systematic analysis of this is clearly very important, but
is beyond the scope of the present work.

\subsubsection{Interacting fermion models}

Our discussion so far has been confined to noninteracting fermions for
which the full hamiltonian is a sum of single-particle
hamiltonians. The generic many-fermion hamiltonian is of the form
\be 
H_F = \sum_{m,n} \frac12 E_{mn} \psid_m \psi_n + 
\sum_{m,n,l,r} V_{mnlr} \psid_{m} \psi_{n}
\psid_{l} \psi_r + \cdots
\label{interacting}
\ee
where $E_{mn}= E_{nm}^*$ and $V_{mnlr},...$ satisfy appropriate
relations to ensure hermiticity of $H_F$ .
By using our bosonization formula \eq{bosonizeapp} we can get
the bosonic version of $H_F$ in terms of the $a, \ad$. (Similarly
we can write down a second bosonized version, in terms
of $b, b^\dagger$ using the results of Section \ref{second}.)
It would be interesting to work out the properties of
these bosons for fermions with a Coulomb interaction, for example,
and compare with the standard collective excitations.

\section{\label{prop}General properties of the bosonized theory}

Let us now explore some consequences of our bosonization. We will
first comment on the finite dimensionality of the single-particle
Hilbert space.

\subsection{Single particle: Quantum phase space is fuzzy and compact}

A finite-dimensional single-particle Hilbert space $\H_N$ is
equivalent to a noncommutative or fuzzy compact phase space
(for reviews on fuzzy spaces, see, e.g. 
\cite{Madore-book,Balachandran:2002ig}).  E.g, if
the $N$ states $\ket i$ are the first $N$ states of a simple harmonic
oscillator:\footnote{We will use $\alpha$, $\alpha^\dagger$ to denote
the lowering/raising operators of the single particle Hilbert space
of a harmonic oscillator (as against $a_i$, $a_i^\dagger$
which are particle creation and annihilation operators).}
\bea
&& \ket i = \frac{\alpha^{\dagger i}}{\sqrt{i!}} \ket 0,
~~ i=0,1,2,\cdots, \, N-1, ~~ \alpha  \ket 0 =0 
\nn
&& \hat h= \hbar (\alpha^\dagger \alpha +\half),~ 
[\alpha, \alpha^\dagger]=1,~\hat h\ket i=
(i+\half)\hbar \ket i \, .
\label{sho}
\eea
then the phase space corresponds to a fuzzy disc (see
e.g. \cite{Lizzi:2003ru} and \eq{fuzzy-disc} below.) 
Similarly, if the $N$ states are the $2j+1
$ states $\ket {j,m}, m=-j, \cdots,+j$ of a rotor then the phase space
is 
a fuzzy sphere \cite{Madore:1991bw}.

The fuzziness or noncommutativity of the quantum phase space ${\cal
M}$ follows from the existence of a finite $\hbar$, irrespective of
whether $N$ is finite or infinite(the infinite-dimensional Hilbert
space of a one-dimensional particle corresponds to a plane with
noncommutative coordinates $[x, p]=i\hbar$). See Section
\ref{husimi.sec} for more.

The compactness of ${\cal M}$ follows from finite $N$. Intuitively,
when the states $\ket i$ are, e.g., ``energy levels'' of a bounded
hamiltonian, the $N$ Bohr orbits occupy a finite area of the phase
space. A more precise construction goes as follows 
\cite{Lizzi:2003ru}. Consider
$\H_N$ as a subspace of an infinite dimensional Hilbert space
$\H_\infty$. For definiteness we will consider the case of the
harmonic oscillator, given by \eq{sho}, the generalization to other
cases being straightforward in principle. The quantum phase space can
be defined in terms of the algebra ${\cal A}_N$ of operators on
$\H_N$, which themselves can be defined from operators on $\H_\infty$
using a projection operator:
\bea
&&\hat O \to \hat O_N \equiv P_N~ \hat O~ P_N= 
\bra i \hat O \ket j \ket i \bra j \in {\cal A}_N,~ i,j=1, \cdots, N 
\nn
&& 
P_N \equiv \sum_1^N \ket i \bra i
\label{trunc}
\eea
Of course, the map $\hat O \to \hat O_N$ is many-to-one. In terms
of the phase space (Husimi) representation of operators (see Appendix
\ref{husimi.sec}), \eq{trunc} reads
\be
{\cal O}(z,\bar z) \to {\cal O}_N(z,\bar z) \equiv 
P_N(z,\bar z)*{\cal O}(z,\bar z)*P_N(z,\bar z)
\label{trunc-husimi}
\ee
where the star product is the Voros star product.
The operation on the RHS essentially turns the support of ${\cal O}(z,\bar z)$ 
into a compact one of an approximate size 
\be
r_0^2 = N\hbar
\label{fuzzy-disc}
\ee
with an exponential tail $\sim \exp[- r^2/\hbar]$. This can be seen by
noting that 
$P_N(z,\bar z)=\Gamma(N+1, {r^2\over 2\hbar})/\Gamma(N+1)$  
has the above fall-off property.

Note that the truncation to an effectively compact phase space does
not depend on taking any semiclassical limit. It is clear that the
support of ${\cal O}_N(z,\bar z)$, irrespective of the original $\hat
O$, will be confined to $r \le r_0$. The geometry of the quantum phase
space is therefore that of a disc. We have proved this
result here for the harmonic oscillator, given by \eq{sho}, but
similar results hold for other finite dimensional Hilbert
spaces.

Another way to see the appearance of the fuzzy disc is to compute the
Husimi distribution for any basis of states in $\H_N$. For the
harmonic oscillator example \eq{sho}, 
using \eq{husimi-sho} we see that the 
Husimi distribution in state $\ket j$ is concentrated around $r =
\sqrt j, j=0,1,2,....,N-1$. The state of the maximum size, with
$j=N-1$, has an approximate radius $r_0$.  The existence of the
maximum size of the Husimi distribution can be easily proved for an
arbitrary linear combination of the basis states by a simple
generalization of the above argument.

\myitem{Coherent states in $\H_N$}

There is another useful basis of states for $\H_N$, the
modified coherent states (see, e.g.
\cite{Pinzul:2001qh}, Eqn. (4.16)) defined using
the projection $P_N$ (see \eq{trunc})
\[
\ket {z,N} = \frac{P_N \ket z}{|| P_N \ket z ||}
\]
which, for the harmonic oscillator example, \eq{sho}, becomes
\bea
\ket {z,N} 
= \frac{1}{( \hbox{Exp}_N[|z|^2])^{1/2}} 
\sum_{k=0}^{N-1} \frac{z^k}{k!} 
\, {(\alpha^{\dagger})}^k \, 
\ket 0
\eea
where $\hbox{Exp}_N [x] \equiv \sum_{k=0}^{N-1} \frac{x^k}{k!}$. The
modified coherent states satisfy completeness relations 
\be
\int d^2 z e^{-|z|^2} \ket{z,N}\bra{ z,N} = P_N.
\ee
The Husimi distribution (regarded in the full phase space of 
$\H_\infty$) of any $\ket{z,N}$ is
\be
H_{z,N}(w,\bar w)=  
\frac{e^{-|w|^2}}{\hbox{Exp}_N [|z|^2]} \, |\hbox{Exp}_N
[\bar w z]|^2.
\ee
which falls off exponentially beyond $|w| = \sqrt{N}$
irrespective of the specific $z$ (it has a
peak around $w \sim z$ if $|z| \le \sqrt N$). 

In terms of the modified coherent states, the Husimi distribution in a state $\ket\psi$, defined in \eq{husimi1-def}), gets modified to
\be
H(z, \bar z, N) =|\langle z, N \ket \psi|^2
\label{husimi-N}
\ee

\myitem{``Fuzzy'' coordinate space}

It is interesting to note that at finite $N$ (and finite $\hbar$) even
the coordinate space is ``fuzzy''.  This is in the sense that, for any
`polarization' $(x,p)$ of the phase space, localization in $x$ ($\Delta
x=0$) requires $\Delta p =\infty$, which is impossible for a compact
phase space. More precisely, the wavefunction $\delta(x- x_0)$ cannot
generically be built by superposing a finite number of wavefunctions
$\chi_i(x)=\langle x| i \rangle, ~ i=1,...,N$. Indeed, the projection
in \eq{trunc} implies that $\ket x$ is replaced by the state $\ket x_N
\equiv P_N \ket x$ which has the following position-space wavefunction
\bea
\overlap y {x_N} = \sum_i 
\overlap y i \overlap i x = 
\sum_{i=1}^N  \chi^*_i(y) \chi_i(x)
\label{x-N}
\eea
This approaches $\delta (x-y)$ only in the limit $N\to \infty$.

\subsection{\label{cake}Second quantization: the bosonic phase space 
density}

Let us define the following second quantized field
\be
\phi(x) \equiv \sum_{i=1}^N a_i \chi_i(x), ~~
\phid(x) \equiv \sum_{i=1}^N \ad_i \chi^*_i(x)  
\label{phi-x}
\ee
The bosonic Fock space state \eq{bose-state} then has the wavefunction
\bea
&& \!\!\!\!\!\!\!\! \overlap{x_1, \cdots, \, x_M}{r_1, \cdots, \, r_N} 
\nn
&& \!\!\!\! = \sum_{\sigma \in P_M} {1\over \sqrt{M!}}
\left[\prod_{i=1}^{r_1}
\chi_{\sigma(1)}(x_i)\prod_{i=r_1+1}^{r_2}\chi_{\sigma(2)}(x_i)
\, \cdots \!\!\!\!\!\!\!\!\!\! \prod_{i=r_1+r_2+ \cdots
+r_{N-1}+1}^{M}\chi_{\sigma(N)}(x_i) 
\right]
\eea
where $\quad \ket{x_1, \cdots, \, x_M}\equiv  
\prod_{l=1}^M \phid(x_l)~\ket{0}, ~~M=r_1 + \cdots + r_N $.
Just like $\ket x$ is outside the single-particle Hilbert space $\H_N$, 
the basis $\ket{x_1, ..., x_M}$
is outside of the Fock space built out of states \eq{bose-state}.
The more appropriate wavefunctions are products of
$\ket x_N$ as in \eq{x-N}. Nevertheless, \eq{phi-x} is
a useful definition to have.

\begin{figure}[ht]
\vspace{0.5cm}
\hspace{-0.5cm}
\centerline{
    \epsfxsize=8cm
   \epsfysize=7cm
   \epsffile{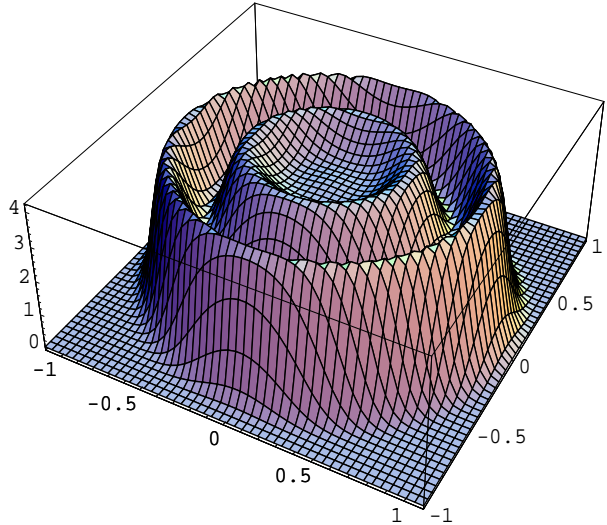}
 }
\caption{\sl  The bosonic phase space density $\vev{W_B(x,p)}$, in the state
\eq{bose-state}, as a function of the $(x,p)$ plane. In the LLM
example, this corresponds to the density of giant gravitons
in the $(x_1,x_2)$ plane.}
\label{heights.fig}
\end{figure}

The second quantized (Wigner) phase space density is given 
in terms of \eq{phi-x} as (cf. \eq{husimi-2b}) 
\be \hat W_B(x,p) = \int
d\eta~e^{i\eta p}~\phi^\dagger (x+ \eta/2)~\phi(x - \eta/2) 
\label{def-ub}
\ee  
The expectation value of $\hat W_B(x,p)$ in the
state $\ket{r_1,..., r_N}$ is easily computed:
\be
\vev{\hat W_B(x,p)} =  \sum_{i=1}^N r_i W_i(x,p),
\label{vev-ub}
\ee
where
\be
W_i (x,p) 
= \int d\eta~e^{i \eta p}~\chi_i(x- \eta/2)~\chi_i^*(x+\eta/2)  
\ee
represents the Wigner density for the individual state $\ket i$.
The semiclassical picture of \eq{vev-ub} for a typical
state is described in Figure \ref{heights.fig}.

For the harmonic oscillator example, one can use \eq{husimi-sho} to
evaluate \eq{vev-ub}. The plot of \eq{vev-ub} in the $(x,p)$ plane
looks like ``a rugged circular cake'' with a maximum diameter $r_0$
given by \eq{fuzzy-disc} and with circular ridges of heights $r_i$ at
radii $\sim \sqrt i$.  For states which are arbitrary linear
combinations of \eq{bose-state} the phase space density still has the
shape of a rugged cake of the same maximum diameter $r_0$, but with
the circular ridges generically replaced by dips and bumps not
necessarily maintaining the circular symmetry.

In the context of LLM, these plots depict the density of giant
gravitons in the $(x_1,x_2)$ plane (see Section \ref{gg.p.d}). 

Another way of seeing the ``rugged cake'' is to compute
\be
\langle z^1_1, ..., z^1_{r_1}, z^2_1,..., z^2_{r_2},
z^N_1,.. , z^N_{r_N} | r_1, ..., r_N \rangle
\sim \hbox{Sym}[\prod_{i=1}^N \prod_{j=1}^{r_i} 
(z^i_j)^i \exp- [\sum_{i=1}^N \sum_{j=1}^{r_i}
 |z^i_j|^2/2]
\ee
which shows that there are $r_k$ particles at radius $\sqrt k$
for each $k=1,\cdots, \, N$. The notation ``Sym'' represents
symmetrization over all the $z$'s. 

A more appropriate second quantized phase space density is
(cf. \eq{husimi-2b})
\bea
&& H_{B,N}(z, \bar z) = \phi(z,\bar z,N) \phid(z, \bar z, N)
\nn
&& \phi(z,\bar z, N) \equiv \sum_{i=1}^N a_i \overlap {z,N}{i}
\label{husimi2-N}
\eea
which are finite $N$ versions of \eq{husimi-2b}. 

\subsection{\label{lattice} Lattice interpretation}

Another, perhaps more appropriate, description of the coordinate space
is in terms of a finite lattice. An easy example is when $\H_N$ is
built out of the first $N$ levels of a harmonic oscillator
(see Eqn. \eq{sho}).  Consider
the radial polarization of the phase space where the angle $\varphi$
is the coordinate and the hamiltonian $r^2/2 = - i \del/\del \varphi$
is the conjugate momentum. The ultra-violet cut-off in energy or the
radius (see \eq{fuzzy-disc}) implies that there is a finite lattice
cutoff in $\varphi$ (this, together with the infra-red cut-off coming
from the compactness of $\varphi$, gives a finite lattice, as
appropriate for a finite dimensional Hilbert space $\H_N$).  Thus
$\varphi$ takes discrete values $\varphi_\mu = \mu \eps,~ \eps= 2\pi / N,
~\mu=0,1,..., N-1, ~ \varphi_\mu \equiv 
\varphi_{\mu+N} = \varphi_\mu + 2\pi$. The associated 
`position eigenstates'
$\ket{\varphi_\mu}$ are defined as discrete Fourier transforms
of the basis states in \eq{sho}:
\be
\ket {\varphi_\mu}
\equiv \sum_{k=1}^N e^{-i {2\pi \over N}\mu k} \ket {k-1}
\ee
The lattice description of the
coordinate space is more appropriate than the
continuum description since with a
continuous variable $\varphi$ the associated states $\ket \varphi$
are typically
outside of $\H_N$. In contrast, $\ket{\varphi_\mu}$ are all
in $\H_N$, by definition.

The second quantized field operator 
$\phi^\dagger (\varphi_\mu) \equiv \phi^\dagger_\mu$, which creates a 
particle on the $\mu$th site, is defined by
\be
\phi^\dagger_\mu \equiv \frac{1}{\sqrt{N}}~
\sum_{k=1}^N e^{-i {2\pi \over N}\mu k} \ad_k
\label{spin} 
\ee
Its conjugate, $\phi_\mu$, destroys a particle on the $\mu$th site. 
It is easy to show that these field operators also satisfy the 
harmonic oscillator algebra
\bea
[\phi_\mu, \phi^\dagger_\nu] &=& 1 \quad {\rm if} \quad 
\mu=\nu \quad {\rm mod} 
\quad N \nn
&=& 0 \quad {\rm otherwise}.
\label{spinalgebra}
\eea
In terms of these field operators, the $N$-level lattice hamiltonian 
may be obtained from \eq{hamiltonian} by using the following relation:
\be
\sum_{i=k}^N \ad_i a_i = \frac{N-k+1}{N}
\sum_{\mu=0}^{N-1} \phi^\dagger_\mu~\phi_\mu+
\frac{1}{N}\sum_{\mu \neq \nu=0}^{N-1} \biggl [\frac{
e^{i {2\pi \over N}(\mu-\nu) k}-e^{i {2\pi \over N}(\mu-\nu)}}
{1-e^{i {2\pi \over N}(\mu-\nu)}} \biggr ]~
\phi^\dagger_\mu~\phi_{\nu} 
\label{transcribe}
\ee 
The hamiltonian is essentially the charge $Q_1$, see equation 
\eq{charges} below. Notice that the second term in the above
equation gives rise to long-range interactions on the lattice. 

Our lattice models inherit higher conserved charges from the
underlying free 
fermion system. There are exactly 
$N$ conserved charges, which arise due to the conservation of the 
individual energies of the non-interacting fermions. It is more 
usual to write these conserved charges as 
\be
Q_n = \sum_{m=0}^\infty \{{\cal E}(m)\}^n~\psid_m \, \psi_m.
\ee
Here $n$ is any positive integer, but the number of independent 
charges is only $N$ since the values of these charges in any 
$N$-fermion state can be expressed in terms of only the $N$ 
independent energies of the occupied levels. In the lattice
formulation of the bosonized theory, these charges translate into
the operators
\be
Q_n \!\! = \!\! \sum_{k=1}^N \biggl \{ {\cal E} \biggl (
N-k + \frac{N-k+1}{N}
\sum_{\mu=0}^{N-1} \phi^\dagger_\mu \, \phi_\mu+
\frac{1}{N}\sum_{\mu \neq \nu=0}^{N-1} \biggl [\frac{
e^{i {2\pi \over N}(\mu-\nu) k}-e^{i {2\pi \over N}(\mu-\nu)}}
{1-e^{i {2\pi \over N}(\mu-\nu)}} \biggr ] \,
\phi^\dagger_\mu \, \phi_{\nu} \biggl ) \biggr \}^n  \nn
\label{charges}
\ee
By construction these higher charges exist for any potential in which
the fermions are moving. In particular, they also exist for fermions
moving in a harmonic oscillator potential, which is relevant for the
half-BPS sector of ${\cal N}=4$ super Yang-Mills theory. This raises
the possibility of some connection of our lattice models with the
integrable spin-chain models of ${\cal N}=4$ super Yang-Mills theory
\footnote{See, for example, the paper by Beisert and Staudacher
\cite{Beisert:2005fw} for a summary of the current status of the
subject and for references to recent literature.} which have been a
recent focus of study in connection with AdS/CFT duality. What this
connection might be is not clear to us, but investigating this
possibility could be worthwhile.

We end this section by mentioning that an explicit expression for the 
lattice 
hamiltonian can be given for an equally spaced spectrum of the form 
${\cal E}(m)=c_1~m+c_2$. This is relevant for the half-BPS sector 
mentioned above. The hamiltonian turns out to be
\be
H_{\rm equal-spacing} = \frac{c_1}{2} (N+1) \sum_{\mu=0}^{N-1} 
\phi^\dagger_\mu~\phi_\mu 
+ \frac{c_1}{2} \sum_{\mu \neq \nu=0}^{N-1} [1-i~\cot{\frac{\pi}{N}(\mu-\nu)}]~
\phi^\dagger_\mu~\phi_{\nu} + H_{\rm vac}
\label{hamiltonian-sho}
\ee 
where $H_{\rm vac}=\frac{c_1}{2}N(N-1)+c_2 N$ is the ground state 
energy. This hamiltonian has long-range interactions. Moreover,
unlike in the case of standard SYM spin-chains 
\cite{Beisert:2005fw}, here there can be 
multiple excitations at any site since the variables at each
site are harmonic oscillators. Whether the reformulation of the
half-BPS sector as the spectrum of this lattice hamiltonian has 
any new insights to offer remains to be seen.

\subsection{More on Second quantization}

In view of the lattice interpretation of coordinate space, it may be
appropriate to introduce discrete Wigner/Husimi phase space distributions
\cite{Chaturvedi:2005ug} $W_B(i,j), H_B(i,j)$ 
in both the first and second quantized formalism (the
precursors of this concept are \eq{husimi-N}, \eq{husimi2-N}). 
We will not go into the full details here, but 
consider only the diagonal elements which turn out
to be
\be
\hat W_B(i,i) = \hat \Phi_B(i,i) =  \ad_i a_i
\label{wb-discrete}
\ee
Using  \eq{bosonizetxt}, we can relate these to $\psid_n \psi_n$
which, when evaluated 
on  wavefunctions  of the form \eq{fermi-state}, get 
related to the fermionic Wigner density 
$\hat W_F(x,p)$ (see \eq{wigner-2}, also footnote \ref{wf-u}), 
\be
\vev{\hat W_F(x,p)}= \sum_n \vev{\psid_n \psi_n} W_n(x,p)
\label{w-circular}
\ee  
Here $W_n$ are the single-particle Wigner distributions.
For the corresponding states \eq{bose-state} we find (see \eq{vev-ub}) 
\be
\vev{\hat W_B(x,p)} =\sum_i \vev{\ad_i a_i} W_i(x,p)
\ee
Combining all this with \eq{bosonizetxt} we get for these special
states (monomials of the form \eq{fermi-state})
\bea
\vev{\hat W_F(x,p)} = 
\sum_n \sum_{k=1}^N \vev{\delta (\sum_{i=k}^N \hat W_B(i,i)
+N -n -k)} W_n(x,p)
\eea
In case of \eq{sho}, $\vev{\hat W_B}, \vev{\hat W_F}$ for these states 
are functions of only $x^2 + p^2$ (``circular configurations'')
in which case we get a relation between the bosonic and
fermionic phase space densities
\bea
\vev{\hat W_F(r)} = 
\sum_n \sum_{k=1}^N
\delta (\sum_{i=k}^N \{\int [dr'] \vev{\hat W_B(r')}
W_i(r') \} +N
-n -k) W_n(r)
\label{uf-ub}
\eea
where we have taken the semiclassical limit to perform the 
average inside the $\delta$-function.

In the next section, we will interpret this as a 
relation between the LLM metric and the giant graviton density.  

\subsection{Action for $W_B$}

For noninteracting fermions, the second quantized action for
the fermions is given by the Kirillov coadjoint orbit action
written in terms of the Wigner phase space
density $\hat W_F$ 
\cite{DMW-nonrel,DMW-path,Mandal:2003tj,Mandal:2005wv,Dhar:2005qh}
\footnote{\label{wf-u}In these references the notation $u(q,p)$
was used in place of $W_F(q,p)$; notations in this paper are
explained in Appendix \ref{husimi.sec}.}
\bea
S[W_F] &=& S_{sympl}[W_F] + S_{ham}[W_F], \cr
S_{sympl} &=& \int dt~ds \int
\frac{dqdp}{2\pi \hbar}~ 
W_F(q,p,t,s)*_w \{\del_t W_F, \del_s W_F\}_{MB}, \cr
S_{ham} &=& \int dt~H_F = \int dt \int 
\frac{dqdp}{2\pi \hbar}~ W_F(q,p,t)*_w h(q,p)
\label{uf-action}
\eea
Here the action is written in the first order form, similar
to the single-particle action
$\int dt~(p\dot q - H(p,q))$, with $W_F(q,p)$ itself playing the role
of the phase space coordinates $(p,q)$. The notation  
``sympl'' denotes the symplectic form ($p\dot q$) term
and ``ham'' denotes the hamiltonian
term. The variable $s$ denotes an extension of $W_F(q,p,t)$ to
$W_F(q,p,t,s)$, devised to enable us to write down the
symplectic term. The physical trajectory is parametrized
by $t$ at some $s=s_0$ and the equation of motion is independent
of the $s$-extension because the symplectic form is closed. 
See \cite{Mandal:2003tj,Mandal:2005wv,Dhar:2005qh} for details.

Using the bosonization formulae in Section 2 and Appendix A 
one gets an action for the bosons
in terms of $W_B(q,p)$. For equally spaced fermionic levels we can
use \eq{uf-ub} to  get
an action of the form
\be 
S[W_B] = S_{sympl}[W_B] + 
\int dt\int \frac{dqdp}{2\pi \hbar}~ W_B(q,p,t) *_w h_B(q,p)
\label{ub-action}
\ee
where $h_B$ is an equally spaced finite-level ($N$) hamiltonian.
The symplectic form can be explicitly written down, though we
will not do so here.

\section{\label{llm}Application of our bosonization to LLM}

\myitem{Review}

As discussed in the Introduction,
\cite{Corley:2001zk,Berenstein:2004kk} made the observation that a
half-BPS sector of ${\cal N}=4$ super Yang Mills theory is described
effectively by a theory of free fermions moving in a simple harmonic
oscillator potential. The semiclassical fermion phase space is
described by droplets of uniform density in two
dimensions. Ref. \cite{LLM} uncovered such a droplet structure also in
the corresponding sector of type IIB supergravity in asymptotically
\ads\ spacetimes. It was noted in
\cite{Corley:2001zk,Berenstein:2004kk, Suryanarayana:2004ig} that
states of the fermion theory should have a bosonic description in
terms of giant gravitons, since the latter are known to correspond to
operators of the Yang Mills theory which can be written in terms of
the fermions \cite{Takayama:2005yq}. The identification
between the fermionic and the bosonic states 
was explicitly stated in \cite{Suryanarayana:2004ig}
as a one-to-one map
\bea
\ket{f_1,\cdots,\, f_N} \leftrightarrow \ket{r_1,\cdots,\,  r_N}  
\eea
where
\bea
&& r_N =  f_1, \nn
&& r_k = f_{N-k+1} - f_{N-k} - 1, \quad \quad k = 1,\,2,\,\cdots\,,
N-1 
\label{statemap} 
\eea
This maps the fermionic hamiltonian
\be
H_F = \sum_{n=0}^\infty (n + 1/2) \psid_n \psi_n,
\ee
with excitation spectrum given by $E=\sum_{k=1}^N (f_k-k+1)$, 
to a bosonic hamiltonian
\be
H_F - N^2/2 = H_B = \sum_{i=1}^N i \ad_i a_i,
\ee
with excitation spectrum given by $E=\sum_{k=1}^N k r_k$. 
A few examples of the map \eq{statemap} are
\bea
\ket{F_0} \equiv \ket{0,1, \cdots, N-1} &&\mapsto \ket{0}_B
\nn
\psid_N \psi_0 \ket{F_0} \equiv \ket{1,2, \cdots, N}
&& \mapsto \ad_N \ket{0}_B
\nn
\psid_N \psid_{N+1} \psi_1 \psi_0 \ket{F_0}
\equiv \ket{2,3,\cdots, N+1}
&& \mapsto\left( (\ad_N)^2/\sqrt{2} \right) \ket{0}_B
\nn
\psi^\dagger_{N+1} 
\psi^\dagger_N \psi_2 \psi_0 \ket{F_0} 
\equiv \ket{1, 3,\cdots, N, N+1}
&& \mapsto \ad_N \ad_{N-1} \ket{0}_B
\label{map-example}
\eea
The second equation assigns a single hole at the lowest level
(costing energy $N$) to a single bosonic particle at level
$N$. The 4th equation assigns two holes at levels 0 and 2
(total energy cost $2N-1$) to two bosonic particles, at
levels $N$ and $N-1$.  

The rationale behind the map \eq{statemap} is as follows.  A fermion
configuration can be specified in terms of holes created in the Fermi
sea. The idea is to regard a hole (together with the upward shift of
the Fermi sea to make space for it) as a bosonic excitation. The
correspondence between giant gravitons and gauge invariant operators
suggests an identification of these bosonic excitations
with giant gravitons.

\myitem{Our operator map and Exactness of the Bose-Fermi equivalence}

The above rationale of treating a hole as a boson is somewhat
intuitive, so let's see how it holds in specific examples.  Consider
the second equation of
\eq{map-example}. Here the hole corresponds to the excitation
$A^\dagger _N \equiv \Phi_{N0} = \psid_N \psi_0$, which satisfies the
algebra (see \eq{winf})
\be
[A_N, A^\dagger_ N]
= \psid_0\psi_0 - \psid_N \psi_N 
\label{naive}
\ee
This evaluates to 1 on $\ket{F_0}$ but not in general. Indeed
these operators are related to the fermion phase space density
which satisfy the \winf\  algebra but not satisfy the Heisenberg algebra.

This might seem to suggest that \eq{statemap} may not hold
operatorially \cite{Berenstein:2005aa}, 
that is, the ``hole'' operators may not
satisfy the usual bosonic commutation rules, after all. On the other
hand, it is also known \cite{Suryanarayana:2004ig}
that  with this map the fermionic and bosonic
partition functions agree. The only way to settle which possibility
is realized is to try to see if one can deduce operator maps from
\eq{statemap}. The operator maps described in this paper (Section 2, 
Appendix A and Section 5) indeed precisely fit the bill
\footnote{We first deduced the operator map for the LLM system,
but as mentioned in Section 2, they hold for fermions moving in an arbitrary
potential and even in
the presence of fermion-fermion interactions}. 
The precise bosonic excitation is more complicated than
the naive guess of \eq{naive} and is such that it does satisfy the
Heisenberg algebra.  Indeed the Heisenberg algebra is implied by the
the fermion anticommutation relations (and vice versa) and the
bose-fermi equivalence is exact.

\subsection{\label{gg.p.d}Giant graviton phase space density} 

The exact bosonic operators $a_i, \ad_i$ are clearly related to
creation or destruction of giant gravitons
\cite{Corley:2001zk,Berenstein:2004kk, LLM,
Suryanarayana:2004ig,Mandal:2005wv}.  To see this in detail, we now
come back to the bosonic phase space discussed in the earlier section.

The correspondences with the last section are: 

1. The states $\ket i \in \H_N$ correspond to giant gravitons
in energy eigenstates.

2. The modified coherent states $\ket{z,N}, |z| < N$
describe a giant graviton state localized near the
point $z$.

3. The giant graviton energy levels are equally spaced, as in the
harmonic oscillator example \eq{sho}. Thus, the phase space density of
giant gravitons ({\it i.e.} the density of giant gravitons in the 
$(x_1, x_2)$
plane)\footnote{The identification of the $(x_1,x_2)$ 
plane of LLM with the phase
space of half-BPS giant gravitons was made in \cite{Mandal:2005wv}.} 
has (see Eqn. \eq{vev-ub} and below) the geometry of a rugged
cake\footnote{There is a subtlety, however, about the origin of the
``cake''. The giant gravitons at the ``North pole'' have zero
energy. If we are not interested in keeping track of the total number
of giant gravitons, we can choose to ignore all such giant gravitons
and therefore ignore the harmonic oscillator ground state. Our $P_N$
(see \eq{trunc}) will therefore consist of the states $\ket i,~
i=1,2,\cdots,\, N$. The formulae in the last section will have to be
correspondingly modified, and in this convention, the rugged cake will
have a ``dip'' in the middle.}, as discussed in section 3.2 (see
Figure \ref{heights.fig}). Such a geometry (with heights $r_i$ at
radii $\sim \sqrt i$) accords with the picture of $r_i$ giant
gravitons moving in the $i$-th orbit. This adds to the evidence that
the bosons $a_i, a_i^\dagger$ indeed represent giant gravitons.

4. Arbitrary LLM `droplet' geometries correspond, in a one-to-one
fashion, to the ``rugged cake'' geometries of giant gravitons. It is
interesting to note that the giant gravitons {\em never leave the
original circular region \eq{fuzzy-disc} representing \ads}, even for
arbitrary LLM geometries.

5. For circular configurations, the giant graviton phase space density
in the semiclassical limit is directly related to the fermion phase
density via \eq{uf-ub}. Because of the relation between the fermion
phase space density and the LLM metric, \eq{uf-ub} also expresses the
LLM metric in terms of the giant graviton phase space density in the
semiclassical limit. Section \ref{llm.g.b} will discuss the finite $N$
version of this relation where we will write down exact expressions
for gravitons in terms of the oscillators $a_i, \ad_j$.

\myitem{Remarks:}

a. The bosonization formula  in Section \ref{bose} automatically
incorporate the fact that although multiple fermions cannot
occupy the same energy levels, the giant gravitons can. 
In the simple examples like \eq{map-example}, where
the RHS of the third line describes two giant gravitons in the $N$-th
orbit, the equivalent fermionic description on the LHS 
describes a spreading out. In general such effects are
encoded in the operator maps.

b. In \cite{Mandal:2005wv}, only non-overlapping giant gravitons were
considered, and agreement found with the $W$-action
\eq{uf-action}. When
overlapping giant gravitons are considered, the data consist of not
only the centres of mass of the giant gravitons, but the ``heights''
(how many on top of each other). It would be an interesting exercise
to obtain \eq{ub-action} from considering such overlapping giant
gravitons. 

\subsection{\label{llm.g.b}LLM gravitons and our bosons}

Consider the single trace operators of the boundary theory. In the
fermionic realization, these operators correspond to 
\cite{Takayama:2005yq}
\be
\beta_m^\dagger = \sum_{n=0}^\infty \sqrt{\frac{(m+n)!}
{2^m n!}}~~
\psid_{n+m}~\psi_n, \quad \quad m=1, \, 2, \, \cdots \, , \infty 
\label{graviton}
\ee 
It is easy to check that $[\beta_m^\dagger,
\beta_n^\dagger]=0=[\beta_m, \beta_n]$, but that $[\beta_m,
\beta_n^\dagger] \neq \delta_{mn}$. In fact, in terms of our bosonic
oscillators which do satisfy the standard oscillator algebra,
\eq{oscillator}, the single trace operators have complicated
expressions, involving creation as well as annihilation operators. The
operator expression, which can be obtained by using the bosonization
formula, \eq{bosonizeapp}, is quite messy, but the corresponding
``single-particle'' state, which is obtained by acting on the fermi
vacuum, has a simple enough expression:
\bea
\beta_m^\dagger~\ket{F_0} = 
&& \sum_{n=2}^N (-1)^{n-1}~\sqrt{\frac{(N+m-n)!}{2^m (m-n)! (N-n)!}}~
\theta_+(m-n)~{\ad_1}^{m-n}~\ad_n~\ket{0}_B \nn
&& ~~~~~~~~~~~~~~~~~~~~~~~~~~~~ 
+\sqrt{\frac{(N+m-1)!}{2^m m! (N-1)!}}~{\ad_1}^m~\ket{0}_B.
\label{singleparticle}
\eea
We see that even a ``single-particle'' state has in general many
excitations of our elementary bosons. For $m<N$, because of the
theta-function the sum in the first term on the right-hand side of
\eq{singleparticle} terminates at $n=m$. So in this case the highest
energy creation operator that appears on the right-hand side is
$\ad_m$ and it appears by itself. If $m<<N$, the leading term on the
right-hand side in $1/N$ expansion has an overall factor of $N^{m/2}$,
which is the correct large-N normalization for a single trace
operator. We thus begin to see how the usual picture of collective
excitations arises in the large-N limit for low-energy states. On the
other hand, for any $m>N$ the highest energy creation operator that
appears in ``single-particle'' states is $\ad_N$ and it appears
together with other excitations. This is a reflection of the fact that
a single trace operator in the boundary theory of higher than $N$th
power of a matrix is not independent since it can be rewritten in terms
of products of lower traces. Let us now explain the above comments in
greater detail.

\subsubsection{\label{fuzzy.sec}Fuzzy gravitons}

{}From \eq{graviton} we get
\bea
\beta^\dagger_1 \ket{F_0} &&= \sqrt{N\over 2} \ad_1\ket 0_B
\nn
\beta^\dagger_2 \ket{F_0} &&= -\half \sqrt{N(N-1)} \ad_2\ket 0_B
+ \half\sqrt{N(N+1)\over 2} (\ad_1)^2 \ket 0_B
\nn
..
\nn
\beta_{N+1}^\dagger~\ket{F_0} =
&& \sum_{n=2}^N (-1)^{n-1}~\sqrt{\frac{(N+m-n)!}{2^m (m-n)! (N-n)!}}~
~{\ad_1}^{N+1-n}~\ad_n~\ket{0}_B \nn
&& ~~~~~~~~~~~~~~~~~~~~~~~~
+\sqrt{\frac{(2N)!}{2^{N+1} (N+1)! (N-1)!}}~{\ad_1}^{N+1}~\ket{0}_B.
\label{fuzzy-gravitons}
\eea
Consider the first equality. Taking  inner product with $\ket z$
on both sides, we see that 
\be
\langle z | \beta^\dagger_1 \ket{F_0} = \sqrt{N\over 2} 
\bra z\ad_1\ket 0_B \sim z \exp[-|z|^2/2]
\ee
This is the wave-function (in the coherent state basis) of the
``graviton'' of unit energy. The ``graviton'' of energy $m \ll N$,
always has a
single particle component $\sim z^m \exp[-|z|^2/2]$. These
correspond to the gravitons wave-functions of \cite{Kim:1985ez}. For
$m>1$, 
however, they also have multi-particle components, e.g. for $m=2$,
there is a two particle component with wavefunction
$\langle z_1, z_2 | (\ad_1)^2  \ket 0_B \sim z_1 z_2 \exp[-(|z_1|^2
+ |z_2|^2)/2]$. For $m>N$ there are no single-particle components of the
wave-functions.

\myitem{Remarks:}

1. It is easy to see from \eq{fuzzy-gravitons} that ``gravitons'' of
energy $>N$ do not exist as single particles and are composed of
multiple $\ad_i$ modes of lower energy. This suggests that the
existence of an infinite number of gravitons, a hallmark of
commutative gravity, is only an approximation valid in the large $N$
limit. For finite $N$, the oscillators $a_i,
\ad_i$, which are by definition $N$ in number and hence independent of each
other, provide a more appropriate basis to describe the geometry.

2. Finite number of independent metric fluctuations is an indication
of noncommutative geometry. Thus, e.g., metric fluctuations on a fuzzy
sphere will involve spherical harmonics only up to rank proportional
to the radius of the fuzzy sphere. 

3. Formulae such as \eq{fuzzy-disc} suggest that 
the large $N$ limit involves nonperturbative effects
of the form  $\exp[-1/\hbar] \sim
\exp[- N]$.

4. At finite $N$, superscript effective action should receive
interesting corrections from the ``compactness'' of the phase space
which is captured by the difference between the Moyal star product and
the ordinary product appearing in the $W_F$ action \eq{uf-action}
\cite{Mandal:2005wv,Dhar:2005qh} and the corresponding structure in
the $W_B$ action \eq{ub-action}.

\section{\label{second}The second bosonization}

Consider a second system of bosons each of which can occupy a state in
an infinite Hilbert space $\H_B$. Suppose we choose a basis $\ket{m},
m=0,\cdots,\infty$ of $\H_B$. In the second quantized notation we
introduce creation (annihilation) operators $b^\dagger_m$ ($b_m$)
which creates (destroys) a particle in the state $\ket{m}$. These
satisfy the commutation relations
\be
\label{sbalgone}
[b_m, b^\dagger_n]= \delta_{mn}, ~~ m, \, n=0,\cdots,\infty
\ee 
A state of this bosonic system is given by (a linear combination
of)
\be
\prod_{k=0}^\infty \frac{(b_k^\dagger)^{n_k}}{\sqrt{n_k!}}
\ket{vac} \, .
\ee
Now consider the subspace of the Hilbert space spanned by states with
the restriction $\sum_{k=0}^\infty n_k = N$. We label a state of this
type by $|s_1, s_2, \cdots, s_N \rangle$ where $s_i$ are
non-increasing set of integers representing the energies of $N$
bosons. In \cite{Suryanarayana:2004ig} a map between these states and
the fermionic states was proposed
\be s_{N-i}
\mapsto f_{i+1} - i, ~~ i = 0,\, 1,\cdots,\, N-1 
\ee 
The rationale behind this map was to regard the ``particle''
excitations (which correspond to dual giant gravitons) as bosons. In
the following subsection we describe the operator equivalence of this
system with the $N$-fermion system.

\subsection{The map between fermions and `dual-giant' bosons}

Let us start by rewriting the Hilbert space of the fermion system into
a disjoint union of subspaces spanned by states with a fixed number of
particle excitations.
\begin{equation}
{\cal H}^{(F)} = \cup_{k=0}^N {\cal H}^{(F)}_k
\end{equation}
where ${\cal H}^{(F)}_{N-k} = \hbox{Span} \{ |0,1,\cdots, k-1,
f_{k+1}, \cdots, f_N \rangle : f_{k+1} \ne k \}$. The operator that
distinguishes the states belonging to different subspaces ${\cal
H}^{(F)}_k$ is $\hat \nu := \sum_{k=1}^\infty b_k^\dagger b_k$ which
translates to
\begin{equation}
\hat \nu = \sum_{k=1}^\infty b_k^\dagger b_k = N - \sum_{k=1}^N  
\prod_{m=0}^{k-1} \psi_m^\dagger \psi_m \, .
\end{equation}
Further, define
\begin{eqnarray}
\hat s_p &=& \sum_{k=1}^\infty (k-1) \sum_{l=0}^{p-1}
\delta(\sum_{i=k}^\infty b^\dagger_i b_i -l) 
\sum_{m=p}^N \delta (\sum_{j=k-1}^\infty b_j^\dagger b_j - m ) , \cr
\hat f_p &=& \sum_{k=0}^\infty k \, \delta ( \sum_{i=0}^{k-1}
\psi_i^\dagger \psi_i -p+1 ) ~ \delta ( \sum_{j=0}^k
\psi_j^\dagger \psi_j -p ).
\label{hat-s-p}
\end{eqnarray}
and
\begin{eqnarray}
\psi_{f_p} &=& \sum_{k=0}^\infty \psi_k ~ \delta (
\sum_{i=0}^{k-1} \psi_i^\dagger \psi_i -p+1 ) ~ \delta (
\sum_{j=0}^k \psi_j^\dagger \psi_j -p ) \cr
b_{s_p} &=& \sum_{k=1}^\infty b_{k-1} \sum_{l=0}^{p-1}
\delta(\sum_{i=k}^\infty b^\dagger_i b_i -l) 
\sum_{m=p}^N \delta (\sum_{j=k-1}^\infty b_j^\dagger b_j - m ) 
\end{eqnarray}
for $p = 1, 2, \cdots, N$. The corresponding $\psi_p^\dagger$ and
$b_{s_p}^\dagger$'s can be obtained by taking the hermitian
conjugates. Since we are to keep the total number of particles fixed
we only consider the operators of the type $b_k^\dagger b_m$. Using
Eqn.(\ref{sbalgone}) we have
\begin{equation}
\label{sbalgtwo}
[b_k^\dagger b_m, ~ b_p^\dagger b_q] = \delta_{mp} b_k^\dagger b_q -
\delta_{kq} b_p^\dagger b_m
\end{equation}
which is again a $W_\infty$-algebra. We want to find the operator
correspondences between operators of the kind $b_k^\dagger b_m$ on the
bosonic side and $\psi_k^\dagger \psi_m$ type operators on the
fermionic side. We present below the expressions of these operators
for the special cases of $N =1, \, 2$. The generalization to
arbitrary $N$ is a lengthy but straightforward exercise.

For $N=1$, we can associate $b_k^\dagger b_m$ with $\psi_k^\dagger
\psi_m$ (and $b_k b_m^\dagger$ with $2\delta_{km} - \psi_k
\psi_m^\dagger$) and the algebra (\ref{sbalgtwo}) follows
immediately.

For $N=2$ case we first seek operators $b_k^\dagger b_0$ ($b_0^\dagger
b_k$) which create (annihilate) a particle excitation that takes a
state in ${\cal H}^{(F)}_m$ to one in ${\cal H}^{(F)}_{m+1}$ (${\cal
H}^{(F)}_{m+1}$ to one in ${\cal H}^{(F)}_m$). To find these let us
observe that each state of the subspace ${\cal H}^{(F)}_k$ is a linear
combination of states with $k$ excited bosons (and $N-k$ in the ground
state) $|s_1, s_2, \cdots, s_k, s_{k+1} = 0, \cdots, s_N = 0 \rangle$
with $k \le N$. The operator $b^\dagger_k b_0$ excites a particle from
level-0 to level-$k$. This operator takes $| s_1=0, \cdots, s_N = 0
\rangle$ to $|s_1 = k, s_2 =0, \cdots, s_N =0 \rangle$. Similarly it
takes $|s_1, 0, \cdots, 0 \rangle$ to $|s_1, k, 0, \cdots 0 \rangle$
if $k \le s_1$ or $|k, s_1, 0, \cdots, 0 \rangle$ if $s_1 \le k$ and
so on.

We find the following expressions for $b_k^\dagger b_0$ and
$b_0^\dagger b_k$:
\begin{eqnarray}
b_k^\dagger b_0 &=& \sqrt{2} \, \psi^\dagger_{k+1} \psi_1\,
\delta_{\hat \nu}
+ \Big[ \sum_{l=1}^{k-1} \psi^\dagger_{k+1} \psi_{0}
\psi^\dagger_{k+l} \, \psi_{k-l+1} \, \delta_{k-{\hat f_2}-l+1} \cr 
&&~~~~~~~~~~~~~~~~~~~~~~~~\, + \psi^\dagger_k \psi_0 \, (
\sum_{l=1}^\infty \delta_{{\hat f_2}-k-l-1} + \sqrt{2} \,
\delta_{k-{\hat f_2}+1})\Big]\, \delta_{{\hat \nu} - 1} , \cr
b_0^\dagger b_m &=& 
\psi_1^\dagger \, \psi_{m+1} \, \delta_{m-{\hat f_2}+1} \,
\delta_{{\hat \nu}-1} +
\Big[ \psi^\dagger_0 \, \psi_{{\hat f_2}} \, \psi_{{\hat f_1}}
  \psi^\dagger_{{\hat f_1}+1}
\, \delta_{{\hat f_2} - m - 1} \, (1-\delta_{{\hat f_2}-{\hat f_1}-1}) 
\cr &&~~~~~~~~~~~~~~~~~~~~~~~~\, + \psi_0^\dagger \, \psi_{m} \,
\delta_{{\hat f_1} -m} \, (1 + \sqrt{2} \, \delta_{{\hat f_2}-{\hat
    f_1}-1}) \Big] \, 
\delta_{{\hat \nu} -2} \,
\end{eqnarray}
for $k,m \ge 1$. The general operators of the type $b_k^\dagger b_m$
can be generated out of the above ones using
\begin{equation}
b_k^\dagger b_m = \delta_{km} \, b_0^\dagger b_0 + [b_k^\dagger b_0,
b_0^\dagger b_m].
\end{equation}
The fermion bilinear operators turn out to be:
\begin{equation}
\psi_k^\dagger \psi_k = (\delta_k + \delta_{k-1})\, \delta_{\hat \nu}
+ (\delta_k + \delta_{k-{\hat s_1}-1}) \, \delta_{{\hat \nu}-1} +
(\delta_{k-{\hat s_2}}  
+ \delta_{k-{\hat s_1}-1}) \, \delta_{{\hat \nu}-2}
\end{equation}
and
\begin{eqnarray}
\psi_{k+1}^\dagger \psi_k &=& \frac{1}{\sqrt{2}} b_1^\dagger b_0 \, 
\delta_{k-1} \, 
\delta_{\hat \nu} + (b_1^\dagger b_0 \, (1-\delta_{{\hat s_1}})
\delta_k + b^\dagger_{k} \, b_{k-1} 
\, \delta_{k-{\hat s_1}-1}) \, \delta_{{\hat \nu}-1} \cr && +
(b^\dagger_{k+1} \, b_{k} \, \delta_{k-{\hat s_2}} \,
(1-\delta_{k-{\hat s_1}}) + b^\dagger_{k} \, 
b_{k-1} \, \delta_{k-{\hat s_1}-1} (1+\frac{1-\sqrt{2}}{\sqrt{2}}
\delta_{k-{\hat s_2}-1})) \, \delta_{{\hat \nu}-2}, \cr  
\psi_{k+n}^\dagger \psi_k &=& \frac{1}{\sqrt{2}}[b_n^\dagger b_0 \,
  \delta_{k-1} + b_{n-1}^\dagger b_1^\dagger b_0^2 \,
 (1+\frac{1-\sqrt{2}}{\sqrt{2}}\delta_{n-2}) \, \delta_k] \,
\delta_{\hat \nu} \cr 
&& + [ b^\dagger_{k+n-1} \, b_{k-1} \, \delta_{k-{\hat s_1}-1} +
  (\sum_{l=2}^\infty b^\dagger_n b_0 \, \delta_{{\hat s_1}-n-l+1}  \cr   
&&
+ \sum_{l=2}^{n-1} b_{n-1}^\dagger \, b_0 \,  b^\dagger_{n-l} \,
b_{n-l-1} \, \delta_{n-{\hat s_1}-l-1} \cr
&& + \frac{1}{\sqrt{2}} ( (b_{n-1}^\dagger)^2 b_0 \,
b_{n-2} \, \delta_{n-{\hat s_1}-2} + \delta_{{\hat s_1}-n} b_n^\dagger 
\, b_0) \, \delta_k] \delta_{{\hat \nu}-1} \cr 
&& + [ b^\dagger_{k+n-1} \, b_{k-1} \, \delta_{k-{\hat s_1}-1} + 
(\sum_{l=2}^\infty b^\dagger_{k+n} \, b_k \, \delta_{{\hat
s_1}-k-n-l+1} \cr && +
\sum_{l=2}^{n-1} b^\dagger_{k+n-1} \, b^\dagger_{k+n-l} \, b_{k+n-l-1} 
\, b_k \,  \delta_{k+n-{\hat s_1}-l-1} \cr 
&& +
\frac{1}{\sqrt{2}} ( b^\dagger_{k+n} \, b_k \, \delta_{{\hat s_1}-k-n}
+ (b^\dagger_{k+n-1})^2 \, b_{k+n-2} \, b_k \, 
\delta_{k+n-{\hat s_1}-2}) \, \delta_{k-{\hat s_2}}] \, \delta_{{\hat
\nu}-2} 
\end{eqnarray}
for $n \ge 2$. The $\delta_{\hat O}$'s in these
expressions are the same as the operator delta functions
$\delta(\hat O)$ used for the first boson
in Section \ref{bose}.
Similar expressions hold for $\psi^\dagger_k \psi_{k+n}$.

The fact that the $N$-fermion system is equivalent to two different
bosonic systems implies that the two bosonic systems are also
equivalent to each other \cite{Suryanarayana:2004ig} (see
\cite{Bena:2004qv} also). As in case of the first boson, one can 
represent various states in this bosonic system by specifying the
corresponding phase space densities. One expects that the phase space
here is the same as that of a harmonic oscillator (i.e, {\bf R}$^2$)
with the total number of particles being equal to $N$. Since they are
bosons, each Planck cell can again be occupied by more than one
particles.

\myitem{Bosonized hamiltonian}

We will consider, for simplicity, bosonized expressions
for non-interacting fermion hamiltonians of the type \eq{free-ham}.
The bosonic hamiltonian, for the $N=2$ system, is
\bea
H_B' \! = \!\!\!  
\sum_{k=0}^\infty {\cal E}(k) [ 
(\delta_k + \delta_{k-1})\, \delta_{\hat \nu}  +
(\delta_k + \delta_{k-{\hat s_1}-1}) \, \delta_{{\hat \nu}-1}  +
(\delta_{k-{\hat s_2}} + \delta_{k-{\hat s_1}-1}) \, \delta_{{\hat
\nu}-2}] - \sum_{k=0}^{N-1} {\cal E}(k)
\nn
\label{hamil-b}
\eea
where ${\hat \nu} = \sum_{n=1}^\infty b^\dagger_n b_n = N-b_0^\dagger
b_0$. The expression for  general $N$ is given in \eq{ham-2}.
Bosonization of \eq{interacting} {\it a la} the second bosonic
system can also be worked out. 

\section{\label{c=1}Applications to $c=1$ matrix model}

In this paper we will only report some preliminary observations on
$c=1$. Detailed results, the full import of which are yet to
be understood, will be presented elsewhere.

As emphasized before, our bosonization formulae do not depend
on the choice of a specific fermion hamiltonian. It can thus
be applied to the $c=1$ model for which the fermion hamiltonian 
involves an upside down harmonic oscillator potential:
\be
H_F = \sum_{n} 
{\cal E}(n) \psid_n \psi_n = \frac12
\int dx~ \psid(x)[-\frac1{\beta^2} \frac{d^2}{dx^2} -
x^2 + A^2]
\psi(x) 
\ee
Here ${\cal E}(n)$ are the eigenvalues of the hamiltonian 
$\hat h = [-\frac1{\beta^2} \frac{d^2}{dx^2} -
x^2 + A^2]$
with an infinite wall at $x = \pm A$ \cite{Moore:1991sf}. 
The spectrum ${\cal E}(n)$ and the corresponding
eigenfunctions can be explicitly evaluated 
in terms of parabolic cylinder functions.
In the scaling limit $N\to \infty, \beta\to\infty,
N/\beta \to A^2/(2\pi)$ one can obtain the following WKB estimate for
energy levels close to the Fermi surface:
\be
{\cal E}(N+m)- {\cal E}(N)
= c m + O(m^2), \, c \equiv  \frac{\pi}{\beta 
\log(\sqrt{2 \beta} A)} 
\label{wkb}
\ee
For the first system of bosons, we get a bosonized
hamiltonian using \eq{hamiltonian}. In the scaling limit
and  for states close to
the fermi level in the fermionic description (corresponding
to bosonic states involving a small number of low energy
bosonic excitations), the hamiltonian \eq{hamiltonian} becomes,
after a simple calculation using \eq{wkb}:
\bea
H_B \approx c \sum_{k=1}^N \sum_{i=k}^N \ad_ia_i
+ \hbox{const.} = c \sum_{k=1}^N k \ad_k a_k + \hbox{const.}
\label{ham-1}
\eea
We will shortly remark on the relation of this
hamiltonian to that of the known relativistic boson.
The hamiltonian in terms of the second system of bosons
can also be written down (generalizing  \eq{hamil-b} to
arbitrary $N$):
\be
H_B' = \sum_{k=1}^N \delta(\sum_{n=1}^\infty b_n^\dagger 
b_n - k) \, \sum_{p=1}^k \left[ {\cal E} 
({\hat s_p} + N -   p) - {\cal E}(N-p) \right] \, . 
\label{ham-2}
\ee
where $\hat s_p$ is defined in \eq{hat-s-p}.

\myitem{Remarks:}

1. {\it D0 branes:}

(a) The bosonic operators $d^\dagger_n \equiv b^\dagger_n b_0$
correspond to creation operators for D0-branes in two dimensional
bosonic string theory (D0 branes in two dimensional string theory were
first described in the matrix model language in
\cite{McGreevy:2003kb,Klebanov:2003km}). This can be verified
as follows. In the fermionic language, $d^\dagger_n$ acting on the
fermi sea kicks the fermion at the fermi level up by $n$ levels; such
an excitation represents a D0 brane at the energy level
$N+n$\footnote{The occasional statements that D0 branes in 2D string
theory correspond to fermions are somewhat loose.  Taken literally,
they would imply that D0 branes cannot be created in a fixed-$N$
theory, at least before the double scaling limit. A more appropriate
picture (emphasized in, e.g.,
\cite{Mandal:2003tj}) is that D0 branes are to be understood as
fermion-antifermion pairs, since kicking a fermion from the fermi
level upwards is such an excitation. Our representation of the D0
brane in terms of the bosonic $d_n, d^\dagger_m$ oscillators is a
precise formulation of this idea.}. Note that the number of
$d^\dagger$ excitations in any state of the bosonized theory is
bounded by $N$, but the individual energy level of such an excitation
is unbounded above (see Section 5), as expected of D0 branes.

(b) A localized D0 brane corresponds to
an appropriate linear combination of $d^\dagger_n$'s so as to form
a single-particle coherent state. Such D0 branes are unstable
and the corresponding ``tachyon potential'' \cite{Sen:1999mg}
is described by the D0 brane hamiltonian \eq{ham-2}. The ``bottom''
of the tachyon potential is, of course, given by the ``vacuum''
state annihilated by all the $d_n, n=1,..,\infty$ oscillators (in 
terms of the $b, b^\dagger$ oscillators this state is
$(b^\dagger_0)^N \ket{vac}, b_n\ket{vac}=0, n=0,1,...,\infty$).
The energy difference $\Delta E$ between the initial energy and the bottom
of the potential can be easily computed using \eq{ham-2}, and for
a D0 brane described by $d^\dagger_k$, $\Delta E= {\cal E}(N+k)
- {\cal E}(N)$.  
 
(c) The hamiltonian \eq{ham-2} describes a nonperturbative interacting
hamiltonian for D0 branes, thus playing the role of a hamiltonian
for open string field theory.

(d) The $d$-oscillators are composites of the $b$-oscillators and do
not have simple Heisenberg commutators. Although $[d^\dagger_m,
d^\dagger_n] =[d_m, d_n]=0,\; [d_n, d^\dagger_m]= b^\dagger_0 b_0 
\delta_{mn} - b^\dagger_m b_n$. It will be interesting if 
this represents interesting statistics among D0 branes.

2. {\it Holes:} The operators $\ad_i$ represent bosonic
excitations corresponding to creation of holes (analogues of the LLM
giant gravitons). In the finite $N$ theory these are again related to
D0 branes, just as for LLM the giant gravitons provide a dual
description of the dual giant states (see Section 5).  It will be
interesting, however, to understand these statements in the double
scaling limit, where these holes are distinguished from the
``particles'' and represent some new states \cite{Sen:2003iv}. This
may provide some nontrivial hints about the double scaling limit
as well as the new states.

3. {\it Tachyons:} It is interesting to understand the double-scaled
limit of our bosonization formulae to find the relation of these
bosons with the tachyon of two-dimensional string theory.  This
problem is similar to the problem of connecting the graviton to the
$a$- and the $b$-oscillators in the LLM case. It turns out that for
small fluctuations around the Fermi vacuum, the hamiltonian \eq{ham-1}
as well as the creation/annihilation operators ($\ad_i, a_j$) coincide
with those for the well-known relativistic 
\cite{Sengupta-Wadia,Das-Jevicki,Polchinski,Gross-Klebanov}
bosons and hence get connected to the massless closed string tachyon
in the double-scaled limit.
Details of this will be presented elsewhere.

\section{Discussion}

We list below the main results and some comments:


1. We have found exact operator bosonizations of a finite number of
fermions which can be described by states in a countable Hilbert
space.

2. There are two systems of bosons. In the first system of bosons,
the number of bosonic particles is not constrained but each bosonic
particle moves in a finite dimensional Hilbert space, the
dimensionality being the number of fermionic particles.  In the second
systems of bosons, the number of bosonic particles has an upper bound
which is the number of fermions and the single-particle Hilbert space
is the same as that of the fermions.

3. In the LLM example, the first kind of bosons correspond to giant
gravitons in the and the second kind of bosons correspond to dual
giant gravitons.

4. In the $c=1$ example, the second kind of bosons correspond to
unstable D0 branes. The interpretation of the first kind of bosons,
which represent holes in the Fermi sea, is not clear.

5. The finite number of bosonic energy 
levels in case of the first bosonization
implies a fuzzy compact phase space (equivalently,
bosons on a lattice). In the LLM case, it has the
important implication that finite rank of the boundary SYM theory
corresponds to NC geometry in the bulk, where the fundamental quanta
describing such geometry are giant gravitons rather than the
perturbative gravitons.

6. The description of $c=1$ in terms of a finite number of bosonic
modes suggests a similar NC structure, although whether it survives 
in the double-scaled limit, and if so in what form, remains an 
open question.

7. The system of fermions on a circle, described in Section
\ref{bose}, is closely related to the problem of formation of 
baby universes as described in \cite{Dijkgraaf:2005bp}. It 
would be interesting to investigate whether in this example also
our bosons are related to microscopic gravitational degrees 
of freedom, as in the LLM case.


It is interesting to speculate whether the NC geometry in the LLM
example is a property only of the specific half-BPS sector of type IIB
theory or if it is more general. One of the points of emphasis in this
paper is that a finite number of modes in the bulk can imply
noncommutative gravity. 
Whether this feature survives in the full theory is an important
question that needs further investigation.

One of the intriguing aspects of our exact bosonization (the first
system of bosons) is that the symmetries of the fermion system (even
the number of spacetime dimensions) are rather intricately hidden in
the bosonic theory. Although at first sight this feature may not
appear to be particularly welcome, it may hint at a more abstract
description of spacetime in which the latter is a derived or emergent
concept. In this context, it might be useful to study the application
of this bosonization to fermion systems in higher dimensions along the
lines briefly outlined in Section 2.1.3.

\gap3

{\bf Acknowledgement:} G.M. would like to thank Perimeter Institute of
Theoretical Physics for a very fruitful stay and especially Jaume
Gomis, Rob Myers and Christian Romelsberger for many discussions
throughout his stay. A.D. would like to acknowledge P.I. for a visit
during the Summer when the collaboration was started.

\gap4

\leftline{\large\bf Appendix}

\gap2

\appendix

\section{\label{details} Details of computations: The first bosonization}

In this Appendix we have put together details of some of the
computations summarized in the main text. We will begin by explaining 
in the first subsection how we arrived at the first bosonization 
described in section 2. In the second  subsection we
will give proof of the algebra \eq{oscillator} for the oscillators
which are defined in \eq{boson1} and \eq{boson2} by their action on arbitrary
fermion states. In the third subsection, we will give an expression
for the general fermion bilinear, $\psid_{n+m}~\psi_n$, 
in terms of the bosonic oscillators and indicate some details of its 
derivation. In the fourth subsection we will prove that this expression
satisfies the \winf\  algebra for small values of $m$ for arbitrary $N$. 
Finally, in the last subsection we will use the expression for the bilinear 
for $N=2$ to prove that it satisfies the \winf\  algebra for arbitrary 
$m$.

\subsection{\label{origin} Origin of the Bosonization formulae}

Here we will describe the steps involved in deducing the bosonization
formulae given in Section 2 from the map \eq{statemap}. Consider first the
action of oscillator creation operators $\ad_k$ for $k < N$ on a
general fermi state. We have,
\bea
\ad_k~\ket{f_1,~\cdots~, f_N}=&& \ad_k~\ket{r_1,~\cdots~, r_k,~\cdots~,r_N} 
\nn = && \sqrt{r_k+1}~\ket{r_1,~\cdots~,r_k+1,~\cdots~, r_N}. \nn
= && \sqrt{f_{N-k+1}-f_{N-k}}~
\ket{f_1,~\cdots~, f_{N-k},~f_{N-k+1}+1,~\cdots~, f_N+1} \nn
\eea
The first and last equalities above follow from the state map
\eq{statemap} and the second follows from the
definition of the creation operator. Similarly, for $\ad_N$ we get
\bea
\ad_N~\ket{f_1,~\cdots~, f_N} = && \ad_N~\ket{r_1,~\cdots~,r_N} \nn
= && \sqrt{r_N+1}~\ket{r_1,~\cdots~, r_N+1}. \nn
= && \sqrt{f_1+1}~\ket{f_1+1,~\cdots~,f_N+1} \nn
\eea
These are exactly the expressions given in equation \eq{boson1}.
Expressions for annihilation operators can be obtained similarly and
these coincide with those given in \eq{boson2}.

Consider now the fermion bilinear $\psid_m~\psi_n$. Let us first set
$m=n$. Acting on a general fermion state, we get
\bea
\psid_n~\psi_n~\ket{f_1,~\cdots~, f_N} 
= && \sum_{k=1}^{N} \delta(f_k-n)~\ket{f_1,~\cdots~, f_N} \nn
= && \sum_{k=1}^{N} \delta \biggl (\sum_{i=1}^k r_{N-k+i}+k-1-n \biggr )~
\ket{r_1,~\cdots~, r_N} \nn
= && \sum_{k=1}^{N} \delta \biggl (\sum_{i=k}^N \ad_i~a_i+N-k-n \biggr )~
\ket{f_1,~\cdots~, f_N}. \nn
\eea
The first equality is a simple consequence of the fact that
$\psid_n~\psi_n$ kills the state unless one of the fermions is
occupying the level $n$. The second and last equalities then follow
from the state map \eq{statemap}. The last
expression coincides with the first line of \eq{bosonizetxt}. The
bosonized formula for the general bilinear given below in \eq{bosonizeapp}
can be obtained by similar manipulations of the state map. The
calculation is essentially elementary, though longer and more tedious.

\subsection{\label{algebraproof} Proof of the oscillator algebra}

Let us consider the first of the equations in \eq{boson1}. Applying
$a_l,~l<k$, on both sides of this equation and using \eq{boson2}, we get 
\bea
a_l \ad_k~\ket{f_1, ..., f_N}  = && \sqrt{f_{N-k+1}-f_{N-k}}~~~a_l~
\ket{f_1, \cdots, f_{N-k}, f_{N-k+1}+1, \cdots, f_N+1} \nn
= && \sqrt{(f_{N-k+1}-f_{N-k})(f_{N-l+1}-f_{N-l}-1)} \times \nn
&& \ket{f_1, \cdots, f_{N-k}, f_{N-k+1}+1, \cdots, f_{N-l}+1, 
f_{N-l+1}, \cdots, f_N}.
\label{a1proof1} 
\eea
Reversing the order, starting with the first of the equations in \eq{boson2} 
and applying a creation operator on both sides, we get
\bea
\ad_k a_l~\ket{f_1, ..., f_N}  = && \sqrt{f_{N-l+1}-f_{N-l}-1}~~~\ad_k~
\ket{f_1, \cdots, f_{N-l}, f_{N-l+1}-1, \cdots, f_N-1} \nn
= && \sqrt{(f_{N-k+1}-f_{N-k})(f_{N-l+1}-f_{N-l}-1)}~ \times \nn
&& \ket{f_1, \cdots, f_{N-k}, f_{N-k+1}+1, \cdots, f_{N-l}+1, 
f_{N-l+1}, \cdots, f_N}. 
\label{a1proof2}
\eea
The right-hand side of this equation is identical to that of \eq{a1proof1}.
It follows that $[a_l, \ad_k]=0$ for $l<k$. One can similarly prove this for
$l>k$. For $l=k$, however, we get
\bea
a_k \ad_k~\ket{f_1, ..., f_N}  = && \sqrt{f_{N-k+1}-f_{N-k}}~~~a_k~
\ket{f_1, \cdots, f_{N-k}, f_{N-k+1}+1, \cdots, f_N+1} \nn
= && (f_{N-k+1}-f_{N-k})~\ket{f_1, \cdots, f_N},
\label{a1proof3} 
\eea
and 
\bea
\ad_k a_k~\ket{f_1, ..., f_N}  = && \sqrt{f_{N-k+1}-f_{N-k}-1}~~~\ad_k~
\ket{f_1, \cdots, f_{N-k}, f_{N-k+1}-1, \cdots, f_N-1} \nn
= && (f_{N-k+1}-f_{N-k}-1)~\ket{f_1, \cdots, f_N}. 
\label{a1proof4}
\eea
If follows that $[a_k, \ad_k]=1$. Combining with the above result, we see 
that our bosonic operators satisfy the standard oscillator algebra 
$[a_l, \ad_k]=\delta_{lk}$.

\subsection{Derivation of the bosonized expression for 
generic fermion bilinear}

We will first give the bosonized expression for the fermion bilinear
and then indicate the key steps in its derivation. The expression given 
below is valid only for $m>0$. The expression for $m<0$ can be 
obtained from it by conjugation. We have,
\bea
&& \psid_{n+m}~\psi_n  \nn
&& = \sum_{k=1}^{N-1} \biggl[\sigma_k^m~
{\sigma_{k+1}^\dagger}^m~\theta_+(\ad_k a_k-m) - \sum_{r_k=0}^\infty~
\sigma_{k-1}^{m-1-r_k}~{\sigma_k^\dagger}^{m-2-r_k}~\sigma_k^{r_k}~
{\sigma_{k+1}^\dagger}^{r_k+1} \nn
&& ~~~~~~~~~\times \theta_-(\ad_k a_k-m+1)~
\theta_+(\ad_{k-1} a_{k-1}+\ad_k a_k-m+1)~
\delta(\ad_k a_k-r_k) \nn
&& ~~~~~~+ \sum_{j=2}^{k-1} (-1)^j \sum_{r_{k-j+1}=0}^\infty 
\sum_{r_{k-j+2}=0}^\infty \cdots \sum_{r_k=0}^\infty 
\sigma_{k-j}^{m-j-\sum_{i=1}^j r_{k-j+i}} \nn  
&& ~~~~~~~~~\times {\sigma_{k-j+1}^\dagger}^{m-j-1-\sum_{i=1}^j r_{k-j+i}}~
\sigma_{k-j+1}^{r_{k-j+1}}~
{\sigma_{k-j+2}^\dagger}^{r_{k-j+1}} \cdots \sigma_{k-1}^{r_{k-1}}~
{\sigma_k^\dagger}^{r_{k-1}} \nn
&& ~~~~~~~~~\times \sigma_k^{r_k}~{\sigma_{k+1}^\dagger}^{r_k+1}~
\theta_- \biggl (\sum_{i=1}^j\ad_{k-j+i}~a_{k-j+i}-m+j \biggr) \nn
&& ~~~~~~~~~\times 
\theta_+ \biggl (\sum_{i=0}^j \ad_{k-j+i}~a_{k-j+i}-m+j \biggr)~
\Pi_{i=1}^j \delta(\ad_{k-j+i}~a_{k-j+i}-r_{k-j+i}) \nn
&& ~~~~~~+ (-1)^k \sum_{r_1=0}^\infty \cdots \sum_{r_k=0}^\infty 
{\sigma_1^\dagger}^{m-1-k-\sum_{i=1}^k r_i}~\sigma_1^{r_1}~
{\sigma_2^\dagger}^{r_1} \cdots \sigma_{k-1}^{r_{k-1}}~
{\sigma_k^\dagger}^{r_{k-1}} \nn 
&& ~~~~~~~~~\times \sigma_k^{r_k}~{\sigma_{k+1}^\dagger}^{r_k+1}~ 
\theta_- \biggl (\sum_{i=1}^k\ad_i a_i-m+k \biggr)~
\Pi_{i=1}^k \delta(\ad_i a_i-r_i) \biggr ] \nn
&& ~~~~~~~~~\times
\delta \biggl (\sum_{i=k+1}^N \ad_i a_i-n+N-k-1 \biggr ) \nn
&& +{\sigma_1^\dagger}^m~
\delta \biggl (\sum_{i=1}^N \ad_i a_i-n+N-1 \biggr ).
\label{bosonizeapp}
\eea 
Let us now explain the main steps in the derivation of this expression.  
Consider the action of the fermion bilinear on a generic state. The result 
is zero unless the level $n$ is occupied, that is
\bea
\psid_{n+m}~\psi_n~\ket{f_1, ..., f_N}  = && \sum_{k=1}^N \delta_{nf_k}~
\psid_{f_1}~\cdots~\psid_{f_{k-1}}~\psid_{n+m}~\psid_{f_{k+1}}~\cdots~
\psid_{f_N} \ket{0}_F \nn
\eea
Furthermore, the right-hand side above vanishes unless the level $(n+m)$ 
is unoccupied. Assuming this is the case, we must consider several 
possibilities, depending on the exact value of $m$. This is done by 
rewriting the above equation as follows:
\bea
&& \psid_{n+m}~\psi_n~\ket{f_1, ..., f_N} \nn
&& = \sum_{k=1}^{N-1} \delta_{nf_k}~ 
\biggl [\sum_{l=f_k+1}^{f_{k+1}-1} \delta_{f_k+m,l}~ 
\psid_{f_1}~\cdots~\psid_{f_{k-1}}~\psid_l~\psid_{f_{k+1}}~\cdots~
\psid_{f_N} \ket{0}_F \nn
&& - \sum_{l=f_{k+1}+1}^{f_{k+2}-1} \delta_{f_k+m,l}~
\psid_{f_1}~\cdots~\psid_{f_{k-1}}~\psid_{f_{k+1}}~\psid_l~
\psid_{f_{k+2}}\cdots~\psid_{f_N} \ket{0}_F \nn
&& + \cdots \cdots \cdots \cdots \cdots \cdots \nn
&& + (-1)^{N-k} \sum_{l=f_N+1}^\infty \delta_{f_k+m,l}~
\psid_{f_1}~\cdots~\psid_{f_{k-1}}~\psid_{f_{k+1}}~\cdots~
\psid_{f_N}~\psid_l~\ket{0}_F \biggr ]\nn 
&& +~~\delta_{nf_N}~  
\psid_{f_1}~ \cdots~\psid_{f_{N-1}}~\psid_{f_N+m}~\ket{0}_F. 
\eea  
The first term in the square brackets above corresponds to the possibility 
that $(n+m)=(f_k+m)$ lies between $(f_k+1)$ and $(f_{k+1}-1)$, the second 
term to the possibility that it lies between $(f_{k+1}+1)$ and $(f_{k+2}-1)$
and so on. The term outside the square brackets corresponds to $k=N$, that is 
to the possibility that $n=f_N$. We can write the above equivalently as
\bea
&& \psid_{n+m}~\psi_n~\ket{f_1,~\cdots~, f_N} \nn
&& = \sum_{k=1}^{N-1} 
\delta_{nf_k}~\biggl [\sum_{l=f_k+1}^{f_{k+1}-1} \delta_{f_k+m,l} 
\ket{\tilde f_1=f_1,~\cdots~, \tilde f_{k-1}=f_{k-1}, \nn
&& \quad \quad \tilde f_k=l, \tilde f_{k+1}=f_{k+1},~\cdots~, 
\tilde f_N=f_N} \nn
&& - \sum_{l=f_{k+1}+1}^{f_{k+2}-1} \delta_{f_k+m,l}~
\ket{\tilde f_1=f_1,~\cdots~, \tilde f_{k-1}=f_{k-1}, \nn
&& \quad \quad \tilde f_k=f_{k+1}, \tilde f_{k+1}=l, 
\tilde f_{k+2}=f_{k+2},~\cdots~, \tilde f_N=f_N} \nn
&& + \cdots \cdots \cdots \cdots \cdots \cdots \nn
&& + (-1)^{N-k} \sum_{l=f_N+1}^\infty \delta_{f_k+m,l}~
\ket{\tilde f_1=f_1,~\cdots~, \tilde f_{k-1}=f_{k-1}, \nn
&& \quad \quad \tilde f_k=f_{k+1}, \tilde f_{k+1}=f_{k+2},~\cdots~,
\tilde f_{N-1}=f_N, \tilde f_N=l} \biggr ] \nn
&& +~~\delta_{nf_N}~\ket{\tilde f_1=f_1,~\cdots~, \tilde f_{k-1}=f_{k-1},   
\tilde f_k=f_{k+1}, \nn
&& \quad \tilde f_{k+1}=f_{k+2},~\cdots~,\tilde f_{N-1}=f_N, 
\tilde f_N=f_N+m}. 
\eea
Using the state map \eq{statemap}, the right-hand side can be 
re-expressed in terms of the bosonic oscillator states and the oscillator
numbers that refer to them. We get,
\bea
&& \psid_{n+m}~\psi_n~\ket{f_1, ..., f_N} \nn
&& = \sum_{k=1}^{N-1} 
\delta \biggl (\sum_{i=1}^k r_{N-k+i}-n+k-1 \biggr )  
\biggl [\theta_+(r_{N-k}-m) \times \nn 
&& \quad \quad \ket{\tilde r_1=r_1,~\cdots~, \tilde r_{N-k-1}=r_{N-k-1},~
\tilde r_{N-k}=r_{N-k}-m, \nn
&& \quad \quad \quad \quad \tilde r_{N-k+1}=r_{N-k+1}+m, \tilde r_{N-k+2}=
r_{N-k+2}~\cdots~, \tilde r_N=r_N} \nn
&& - \theta_-(r_{N-k}-m+1)~\theta_+(r_{N-k-1}+r_{N-k}-m+1) \times \nn
&& \quad \ket{\tilde r_1=r_1,~\cdots~, \tilde r_{N-k-2}=r_{N-k-2},
\tilde r_{N-k-1}=r_{N-k-1}+r_{N-k}-m+1, \nn
&& \quad \quad \tilde r_{N-k}=m-2-r_{N-k}, 
\tilde r_{N-k+1}=r_{N-k+1}+r_{N-k}+1, \nn 
&& \quad \quad \quad \tilde r_{N-k+2}=r_{N-k+2}~\cdots~, 
\tilde r_N=r_N} \nn
&& + \cdots \cdots \cdots \cdots \cdots \cdots \nn
&& + (-1)^{N-k} \theta_-\biggl ( \sum_{i=1}^{N-k} r_i-m+N-k \biggr ) 
\times \nn
&& \quad \ket{\tilde r_1=\sum_{i=1}^{N-k} r_i-m+N-k,~
\tilde r_2=r_1,~\cdots~, \tilde r_{N-k}=r_{N-k-1}, \nn
&& \quad \quad \tilde r_{N-k+1}=r_{N-k+1}+r_{N-k}+1, 
\tilde r_{N-k+2}=r_{N-k+2},~\cdots~,\tilde r_N=r_N} \biggr ] \nn
&& + \delta \biggl (\sum_{i=1}^N r_i-n+N-1 \biggr )~
\ket{\tilde r_1=r_1+m,~\tilde r_2=r_2,~\cdots~, \tilde r_N=r_N}. 
\eea
Now, using the bosonic creation and annihilation operators it is easy 
to re-express every bosonic state appearing on the right-hand side above in 
terms of the state $\ket{r_1,~\cdots~, r_N}$ to which the fermionic state
$\ket{f_1,~\cdots~, f_N}$ corresponds under the state map \eq{statemap}. 
This results in the bosonized operator expression for the fermi bilinear
which is given in \eq{bosonizeapp}. 

\subsection{Proof of \winf\  algebra for small values of $m$ for 
arbitrary $N$}

Consider the bilinears $\psid_{n+1}~\psi_n$ and $\psid_{n+2}~\psi_{n+1}$. 
{}From the \winf\  algebra \eq{winf}, we get
\be
[\psid_{n+2}~\psi_{n+1}, \psid_{n+1}~\psi_n]=\psid_{n+2}~\psi_n.
\ee
It is clear from this equation that we can generate bosonized expressions 
for $\psid_{n+m}~\psi_n$ for any $m$ merely from the knowledge of 
bosonized expression for $\psid_{n+1}~\psi_n$ by using the \winf\  algebra. 
However, here we will use the expressions given in \eq{bosonizetxt}
for $m=0,~1$ and $2$, 
which are special cases of \eq{bosonizeapp}, to compute the commutator 
and verify that the result agrees with the right-hand side. 

To compute the commutator we will need to use the following identities,
in addition to those given in \eq{sigmarelations}:
\be
g(\ad_k a_k)~\sigma_k^\dagger=\sigma_k^\dagger~g(\ad_k a_k+1), \quad 
g(\ad_k a_k)~\sigma_k=\sigma_k~g(\ad_k a_k-1)~\theta_+(\ad_k a_k-1),
\label{sigmarelationsapp}
\ee 
where $g$ is any function of the number operator. We are now ready to 
do the computation using the second of \eq{bosonizetxt} in the commutator.
All terms in the commutator involve products of two delta-functions 
and several vanish because these are incompatible. The surviving
terms are
\bea
&& [\psid_{n+2}~\psi_{n+1}, \psid_{n+1}~\psi_n] \nn
&& = \sigma_1^\dagger~\delta \biggl (\sum_{i=1}^N
\ad_i a_i-n+N-2 \biggr)~\sigma_1^\dagger~\delta \biggl (\sum_{i=1}^N
\ad_i a_i-n+N-1 \biggr) \nn
&& + \sigma_1^\dagger~\delta \biggl (\sum_{i=1}^N \ad_i a_i-n+N-2 \biggr)~
\sigma_1~\sigma_2^\dagger~\theta_+(\ad_1 a_1-1)~
\delta \biggl (\sum_{i=2}^N \ad_i a_i-n+N-2 \biggr) \nn
&& - \sigma_1~\sigma_2^\dagger~\theta_+(\ad_1 a_1-1)~
\delta \biggl (\sum_{i=2}^N \ad_i a_i-n+N-2 \biggr)~
\sigma_1^\dagger~\delta \biggl (\sum_{i=1}^N \ad_i a_i-n+N-2 \biggr) \nn
&& + \sum_{k=1}^{N-1} \sigma_k~\sigma_{k+1}^\dagger~
\theta_+(\ad_k a_k-1)~\delta \biggl (\sum_{i=k+1}^N \ad_i a_i-n+N-k-2 \biggr)
\nn
&& \times~
\sigma_k~\sigma_{k+1}^\dagger~
\theta_+(\ad_k a_k-1)~\delta \biggl (\sum_{i=k+1}^N \ad_i a_i-n+N-k-1 \biggr)
\nn
&& - \sum_{k=2}^{N-1} \sigma_k~\sigma_{k+1}^\dagger~
\theta_+(\ad_k a_k-1)~\delta \biggl (\sum_{i=k+1}^N \ad_i a_i-n+N-k-1 \biggr)
\nn
&& \times~
\sigma_{k-1}~\sigma_k^\dagger~\theta_+(\ad_{k-1} a_{k-1}-1)~
\delta \biggl (\sum_{i=k}^N \ad_i a_i-n+N-k-1 \biggr) 
\eea
The first term comes from the commutator of the first term in the 
bosonized expression for the bilinear. The next two terms come from 
the cross-commutator between the first term and $k=1$ piece of the 
second term (which involves sum over $k$). The last two terms are from 
the commutator of the second term; this survives only when the same 
$k$ is picked from the two sums or if the $k$'s differ by $1$. Further 
simplification requires the use of the relations \eq{sigmarelationsapp}, 
\eq{sigmarelations} and $\theta_+(\ad a-1)=\theta_+(\ad a)-\delta(\ad a)=
1-\delta(\ad a)$. The result is precisely the expression on the 
right-hand side of the of the last of \eq{bosonizetxt}.

Another test of the \winf\  algebra comes from the use of the bosonized
expression of the fermion bilinear conjugate to $\psid_{n+1}~\psi_n$. 
We have,
\bea
\psid_n~\psi_{n+1} = && {\biggl (\psid_{n+1}~\psi_n \biggr )}^\dagger 
= \sigma_1~\delta \biggl (\sum_{i=1}^N \ad_i a_i-n+N-2 \biggr) \nn
&& + \sum_{k=1}^{N-1} \sigma_{k+1}~\sigma^\dagger_k~\delta \biggl
(\sum_{i=k+1}^N \ad_i a_i-n+N-k-2 \biggr).
\eea
{}From the \winf\  algebra we see that
\be
[\psid_{n+1}~\psi_n, \psid_l~\psi_{l+1}] = \delta_{nl}
(\psid_{n+1}~\psi_{n+1}-\psid_n~\psi_n).
\label{conjugate}
\ee
Using the bosonized expressions in the commutator, we get
\bea
&& [\psid_{n+1}~\psi_n, \psid_l~\psi_{l+1}] \nn
&& = \delta_{nl} \biggl \{ \theta_+(\ad_1 a_1-1)~\delta \biggl
(\sum_{i=1}^N \ad_i a_i-n+N-2 \biggr)-\delta \biggl
(\sum_{i=1}^N \ad_i a_i-n+N-1 \biggr) \nn
&& + \sum_{k=1}^{N-1} \biggl [\theta_+(\ad_{k+1} a_{k+1}-1)~
\delta \biggl (\sum_{i=k+1}^N \ad_i a_i-n+N-k-2 \biggr ) \nn
&& - \theta_+(\ad_k a_k-1)~
\delta \biggl (\sum_{i=k+1}^N \ad_i a_i-n+N-k-1 \biggr ) 
\biggr ] \biggr \} 
\eea
Now, let us replace $\theta_+(\ad a-1)$ by the equivalent expression
$(1-\delta(\ad a))$ in all the three places. All the terms containing 
double delta-functions mutually cancel, except the one coming from $k=(N-1)$
of the first term in square brackets. But this has two incompatible     
delta-functions and so vanishes. The result for the right-hand side is
\be
\delta_{nl} \sum_{k=1}^N \biggl [\delta \biggl
(\sum_{i=1}^N \ad_i a_i-n+N-k-1 \biggr)-\delta \biggl
(\sum_{i=1}^N \ad_i a_i-n+N-k \biggr) \biggr ],
\ee 
which is precisely the bosonized expression one gets by using the first
of \eq{bosonizetxt} in the the right-hand side of \eq{conjugate}. We see
that for this test to work out, delicate cancellations between various  
terms were required. 

\subsection{Proof of \winf\  algebra for all $m$ for $N=2$}

For $N=2$, bosonized expressions for the bilinear for all values of 
$m$ have been given in \eq{bosonizeN2}. This is the first nontrivial, 
yet calculationally manageable case. We have checked that in this case 
the \winf\  algebra is satisfied. Here we will indicate the main steps 
in the calculation. Like in the above calculations, delicate cancellations 
between various terms in the commutator are required for the algebra 
to work out, as we shall see.

We will be interested in the commutator
\be
[\psid_{n+m}~\psi_n, \psid_{l+p}~\psi_l] =
\delta_{l+p,n}~\psid_{l+p+m}~\psi_l - \delta_{n+m,l}~\psid_{n+m+p}~\psi_n,  
\label{algebra}
\ee
for $m \geq p>0$. Other cases can be treated similarly. 

There are three terms in the bosonized expression for the fermion 
bilinear given in \eq{bosonizeN2}. It is easy to see that the 
self-commutator of the first two terms reproduces the first two terms
required by the bosonized expression for the right-hand side of  
\eq{algebra}. To prove that the bosonized expressions satisfy the
\eq{algebra}, we then need to show that the self-commutator of the third 
term, together with all the cross-commutator terms, reproduces the 
required third term on the right-hand side. The self-commutator of 
the third term works out to be
\bea
&& {\sigma_1^\dagger}^{m-p}~{\sigma_2^\dagger}^p~
\theta_+(p+l-n-1)~\delta(\ad_1 a_1-n+l+1)~\delta(\ad_2 a_2-l) \nn
&& - \sigma_1^{m-p}~{\sigma_2^\dagger}^m~\theta_+(p+l-m-n-1)~
\theta_+(m+n-l-1) \nn
&& \times \delta(\ad_1 a_1-l+n+1)~\delta(\ad_2 a_2-n),
\label{commutator1}
\eea
while the cross-commutator of the first two terms gives
\bea
&& \sigma_1^{m-p}~{\sigma_2^\dagger}^m~[\theta_+(p+l-m-n-1)-
\theta_+(l-n-m-1)] \nn
&& \times \delta(\ad_1 a_1-l+n+1)~\delta(\ad_2 a_2-n) \nn
&& - {\sigma_1^\dagger}^{m-p}~{\sigma_2^\dagger}^p~\theta_-(n-p-l-1)~
\delta(\ad_1 a_1-n+l+1)~\delta(\ad_2 a_2-l).
\label{commutator2}
\eea
The contributions in \eq{commutator1} and \eq{commutator2} are very 
similar and can be combined. Thus the first term of \eq{commutator1}
can be combined with the second term of \eq{commutator2} using 
$\theta_+(p+l-n-1)-\theta_-(n-p-l-1)=-\delta_{p+l,n}$. Simplifying in 
this way gives the net combined contribution 
\bea
&& - \delta_{p+l,n}~{\sigma_1^\dagger}^{m-p}~{\sigma_2^\dagger}^p~
\delta(\ad_1 a_1-p+1)~\delta(\ad_2 a_2-l) \nn
&& + \delta_{m+n,l}~\sigma_1^{m-p}~{\sigma_2^\dagger}^m~
\delta(\ad_1 a_1-m+1)~\delta(\ad_2 a_2-n),
\label{commutator3}
\eea
Finally, the cross-commutator between the first two terms 
and the third term gives
\bea
&& \biggl [ - \delta_{p+l,n} \sum_{r_1=0}^{p-2} 
{\sigma^\dagger_1}^{m+p-2-r_1}~\sigma_1^{r_1}~
{\sigma_2^\dagger}^{r_1+1}~\delta (\ad_1 a_1-r_1 )~\delta (\ad_2 a_2-l) \nn
&& + \delta_{m+n,l} \sum_{r_1=0}^{p-2} 
{\sigma^\dagger_1}^{p-2-r_1}~\sigma_1^{r_1+m}~
{\sigma_2^\dagger}^{r_1+m+1}~\delta (\ad_1 a_1-r_1-m )~\delta (\ad_2 a_2-n)
\biggr ] \nn
&& - \biggl [ m \leftrightarrow p, n \leftrightarrow l \biggr ].
\eea
Notice that by changing the summation variable from $r_1$ to 
$(r_1+m)$ in the second term in the first square brackets, we 
get a summand that is similar to the first term, but with a 
summation range for the new variable from $m$ to $(m+p-2)$.
This nicely combines with the first term in the second square
brackets to give an overall summation range for $r_1$  from  
$0$ to $(m+p-2)$, as is required if \winf\  algebra is to be 
satisfied by the bosonized expressions. To be precise, 
the summation range in the first term in the second square
brackets is from $0$ to $(m-2)$ only, so the contribution 
for $r_1=(m-1)$ is missing from the extended summation range.
Fortunately, the terms in \eq{commutator3} precisely supply 
this missing contribution. Taking this into account and simplifying, 
we find that the net result of the commutator calculation is 
\bea
&& - \delta_{p+l,n} \sum_{r_1=0}^{m+p-2} 
{\sigma^\dagger_1}^{m+p-2-r_1}~\sigma_1^{r_1}~
{\sigma_2^\dagger}^{r_1+1}~\delta (\ad_1 a_1-r_1 )~\delta (\ad_2 a_2-l) \nn
&& + \delta_{m+n,l} \sum_{r_1=0}^{m+p-2} 
{\sigma^\dagger_1}^{m+p-2-r_1}~\sigma_1^{r_1}~
{\sigma_2^\dagger}^{r_1+1}~\delta (\ad_1 a_1-r_1)~\delta (\ad_2 a_2-n).
\eea
This is exactly the third term in the bosonized form of the 
right-hand side of \eq{algebra}. Hence we have proved that our 
bosonization satisfies the \winf\  algebra for $N=2$. 

\section{\label{husimi.sec} Quantum phase space distributions and star
products}

Some assorted references carrying more detailed versions of formulae
in this Appendix are
\cite{DMW-classical,Madore-book,Balachandran:2002ig,Iso:1994ze,Lizzi:2003ru,Pinzul:2001my,Pinzul:2002fi,Chaturvedi:2005ug,Balasubramanian:2005mg}.

\subsection{Single particle}

Consider the  (infinite-dimensional) Hilbert space of a
particle in one dimension, carrying a representation
of the Heisenberg algebra 
$[\hat x, \hat p] = i\hbar$. The 
Wigner phase space  distribution of the particle 
in a wavefunction $\ket \psi$ is given by
\bea
W(q,p) \equiv 
\int d\eta~ e^{-i p \eta/\hbar}~\psid(q - \eta/2)~\psi(q + \eta/2)
= \bra \psi \hat g_{q,p} \ket \psi, ~\hat g_{q,p} \equiv
e^{i\frac{q \hat p + p \hat x}{\hbar}} 
\label{wigner1-def}
\eea 
There are other choices for the phase space distribution,
e.g. the Husimi distribution,  given by
\bea
&& H(z, \bar z) \equiv  | \langle z | \psi \rangle |^2
= \int{ dq'dp'\over 2\pi\hbar}
W(q',p') e^{-\frac1{2\hbar}[(q'- q)^2 + (p'-p)^2]},~
z \equiv q + ip 
\label{husimi1-def}
\eea
Using Wigner phase space distribution one can define a
map between  operators and functions, with an associated star product:
\bea 
&& \hat O \mapsto O_w(q,p) \equiv {\rm Tr} (\hat O g_{q,p}),~
\bra \psi \hat O \ket \psi = \int dq dp O_w(q,p) W(q,p)
\nn
&& \hat A \hat B \mapsto A_w(q,p) *_w B_w(q,p) \equiv
e^{i\hbar[ \del_p \del_{q'} - \del_{p'} \del_q]} (A_w(q,p) B_w(q',p'))|{q'=q,
p'=p} 
\nn
\eea
The inverse map $O_w \to \hat O$ corresponds to Weyl operator
ordering of the function $O_w$.
The above star product $*_w$ is called the Moyal star product. 

The corresponding definitions for Husimi distribution are
\bea 
&& \hat O \mapsto O_h(z,\bar z) \equiv \bra z \hat O \ket z,~
\bra \psi \hat O \ket \psi = \int dq dp O_h(q,p) H(q,p)
\nn
&& \hat A \hat B \mapsto A_h(z,\bar z) *_h B_h(z, \bar z) \equiv
e^{i\hbar[ \del_z \del_{\bar z'} - 
\del_{z'} \del_{\bar z}]} (A_w(z,\bar z) B_w(z',\bar z'))|{z'=z}
\eea
where $*_h$ is called the Voros star product.

Example of the harmonic oscillator \eq{sho}:

The Wigner and Husimi distributions  for the wavefunction $\ket j,
j=0,1,...,\infty$, are
\bea
&& W_j(q,p) = {(-1)^j\over \pi\hbar} 
e^{-\frac{q^2 + p^2}{\hbar}} L_j(\frac{2(p^2 + q^2)}{\hbar})
\nn
&& 
H_j(z, \bar z) = \frac1{2\pi\hbar (j!)} e^{-|z|^2} |z|^{2j}
\label{husimi-sho}
\eea 

\subsection{Second quantization}

\myitem{Fermions:}

Consider a system of $N$ fermions, as in Section 2.

The operators $\Phi_{mn} = \psid_m \psi_n$ (see Eqn. \eq{phi-mn})) 
satisfy the \winf\ algebra [DMW]
\be
[\Phi_{mn}, \Phi_{m'n'}] = \delta_{m'n}\Phi_{mn'} - \delta_{mn'}\Phi_{m'n}
\label{winf}
\ee
$\Phi_{mn}$ are the basic operators in any
one dimensional fermion field theory in a given fermion number
sector. 
A basis free notation is
\be
\Phi = \sum_{m,n} \Phi_{mn} \ket n \bra m 
\ee
The second quantized Wigner phase space density $\hat W_F(q,p)$
is a linear combination of $\Phi_{mn}$ 
\bea
\hat W_F(q,p) = Tr (\Phi g_{q,p})
= \int d\eta~ e^{-i p \eta/\hbar}~\psid(q - \eta/2)~\psi(q + \eta/2)
\label{wigner-2}
\eea  
The expectation value of $\hat W_F(q,p)$ in the fermi state
\eq{fermi-state} is the sum of the single-particle distributions, 
$\sum_m W_{f_m}(q,p)$. 
The second quantized Husimi phase space density $\hat H_F(q,p)$
is given by
\bea
\hat H_F(z, \bar z) = Tr (\Phi \ket z\bra z) 
= \sum_{m,n} \psid_m \psi_n (\chi_m(z))^* \chi_n(z),~~
\chi_n(z) = \overlap z n
\label{husimi-2}
\eea 

\myitem{Bosons:}

The second quantized phase space distributions for bosons are given by
similar formulas, 
\bea
&& 
\hat W_B(q,p) = Tr (\Phi_B g_{q,p})
= \int d\eta~ e^{-i p \eta/\hbar}~\phid(q - \eta/2)~\phi(q + \eta/2),
~ \Phi_B = \phid_i \phi_j \ket i \bra j
\nn
&& \hat H_B(z, \bar z) = Tr (\Phi \ket z \bra z )
= \sum_{i,j} \phid_i \phi_j (\chi_i(z))^* \chi_j(z),~~
\chi_i(z) = \overlap z i
\label{husimi-2b}
\eea 

\section{\label{rep}New bosonic oscillator representation of $U(K)$}

We begin by noting that the $W_\infty$ algebra \eq{winf}, generated by
$ \psid_m \psi_n, m,n=0,1,...\infty$, has the following nested subalgebras
\be
U(1) \subset U(2) \subset U(3) ... \subset W_\infty
\label{nested}
\ee 
where the subalgebra $U(K), K=1,...,\infty$ is generated by
the finite $\psid_m \psi_n, m,n=0,1,...,K-1$. The structure
constants in \eq{winf} are easily seen to be the structure 
constants of $U(K)$.

The representation of the subalgebra $U(K)$, provided by $\F^K_N$,
defined as the Hilbert space of $N$ fermions in the first $K$ levels
$m=0,1,...,K-1$, is the rank-$N$ antisymmetric tensor representation (dimension
$^KC_N$).

We will bosonize $\F^K_N$ and its operators,
using \eq{bosonizeapp} and its special cases and in
the process will obtain 
novel\footnote{different from Schwinger representations where the generators
are bilinears in bosonic oscillators.} bosonic representations of $U(K)$. 

We will start with the simplest examples of small $N$.

\subsection{The $N=1$ example} 

Here $\F^K_N = \H^K$, the single-particle Hilbert space
of fermions, truncated to the first $K$ levels.
We rewrite the equations of \eq{bosonizeN1} involving 
the first $K$ of fermionic oscillators. For
$m,n=0,1,...,K-1$,
\bea
\psid_{m}\psi_m &&=  P_m \equiv \delta(\ad a - m)
\nn
\psid_{m}\psi_{n} &&=   (\sigma^\dagger)^{m-n} P_n,~m > n
\nn
\psid_{m}\psi_{n}  &&=  P_m (\sigma)^{n-m}, ~m < n 
\label{n=1bose}
\eea
Here $P_m = \ket m \bra m$, the projection operator. $a, \ad$
denote $a_1, \ad_1$ and $\sigma, \sigma^\dagger$ are defined
as in \eq{sigmas}.

Now, although $a, \ad$, and consequently $\sigma, \sigma^\dagger$,
are infinite dimensional matrices (Heisenberg algebra can 
only have infinite dimensional representations), the operators
on the RHS of \eq{n=1bose} have  the matrix form (in the
basis ${(\ad)^m\over \sqrt{m!}}\ket 0$)
\be
\mattwo{A}000
\label{block}
\ee
where $A$ is a $K \times K$ matrix. 

Since the operator map \eq{bosonizeN1} ensures that algebra of fermion
bilinears is reproduced by the bosonic expressions, the right hand
side of \eq{n=1bose} provides a bosonic representation of $U(K)$. We
will consider some explicit, small $K$, examples below.

\myitem{The case $K=2$: bosonic representation of U(2) or SU(2)}

For $K=2$ $\Phi_{mn}$ generate the $U(2)$ algebra. The bosonic versions
of the generators are $P_0, P_1, \ad P_0, P_0 a$. These correspond to
matrices of the form \eq{block}, where $A$ is a $2 \times 2$ matrix,
assuming the following values, respectively
\be
P_0 \to \mattwo1000,  P_1 \to \mattwo0001,
\ad P_0  \to \mattwo0100,  P_0 a\to \mattwo0010
\ee
These provide a bosonic construction of the spin-1/2 representation
of SU(2) ($P_0 + P_1$ represents the trace part of $U(2)$ algebra). 

\myitem{$K=3$: bosonic representation of U(3) or SU(3)}

\eq{n=1bose} now gives the following bosonic generators of $U(3)$:
 $P_0, P_1, P_2, \ad P_0, {1\over \sqrt 2}(\ad) P_1, (\ad)^2 P_0, P_0
 a, {1\over \sqrt 2} P_1 a, P_0 a^2$.  $P_0 + P_1 + P_2$ represents
 the trace part and the rest provide the fundamental representation ${\bf 3}$ of
 $SU(3)$. The matrix representations are of the form \eq{block} with
 $A$ equal to standard $3 \times 3$ $SU(3)$ matrices.

For general $K$ ($N=1$) we get a bosonic construction of the
fundamental (dim $K$) representation of $SU(K)$.

\subsection{The $N=2$ example}

The relevant bosonization formulae are \eq{bosonizeN2}. The bosonic
Hilbert space is a linear combination of states $\ket{mn} = {(\ad_1)^m
(\ad_2)^n \over \sqrt{m! n!}} \ket{00}$.  Let us define projectors
$P_{mn} =\ket {mn} \bra {mn}= \delta(\ad_1 a_1 -m) \delta(\ad_2 a_2
-n)$.

We will start with examples of small $K$.
The first non-trivial case is $K=3$, for which \eq{bosonizeN2} gives
\begin{eqnarray*}
&& \psid_0 \psi_0 = \sum_{m=0}^\infty P_{m0},
\psid_1 \psi_1 = P_{00} + \sum_{m=1}^\infty P_{m1},
\psid_1 \psi_1 = P_{01} + P_{10} + \sum_{m=0}^\infty P_{m2}
\nn
&& \psid_1 \psi_0 = \sigma_1 \sigma^\dagger_2 \sum_{m=0}^\infty P_{m0},
\psid_2 \psi_0 =(\sigma_1 \sigma^\dagger_2)^2 
\sum_{m=0}^\infty P_{2m} -\sigma^\dagger_2P_{00},
\psid_2 \psi_1 =\sigma_1^\dagger P_{00} +
\sigma_1 \sigma^\dagger_2 \sum_{m=1}^\infty P_{m1}
\end{eqnarray*}
The bosonic operators are infinite
dimensional matrices, but are of a triangular form (cf. \eq{block})
\be
\mattwo A{A'} 0 {A''}
\ee
where $A$ is a $3 \times 3$ matrix, corresponding to the
subspace $\H_3= Span\{\ket {00}, \ket {01}, \ket{10}\}$. 
The matrices $A$ can be worked out and they correspond to
an irrep. of $SU(3)$ (viz. the representation $\bar {\bf 3}$).  
  
\subsection{\label{lemma}Bosonization of $N$ fermions in a $K$-level system}

We will now give the result for general $N, K$ which is
straightforward to derive:

\begin{itemize}

\item The bosonization formulae \eq{bosonizeapp} can be applied to
bosonize $N$ fermions in a $K$-level system.

\item The bosonization formulae \eq{bosonizeapp} 
give a novel bosonic construction of the
general rank-$N$ antisymmetric tensor rep of $SU(K)$ in terms of $N$ bosonic
oscillators.

\end{itemize}

\bibliographystyle{utphys} 
\bibliography{myrefs}

\end{document}